\documentclass[12pt]{article}
\usepackage{latexsym,epsfig,amssymb, amsmath, cite, mcite}
\numberwithin{equation}{section}

\usepackage{amsmath}
\usepackage{amsfonts}
\usepackage{dsfont}
\usepackage{amssymb, multirow}
\usepackage{pdfsync}
\usepackage{epsfig}
\usepackage{graphicx}
\usepackage{subfigure}
\usepackage{nicefrac}

\newcommand{\be}{\begin{equation}}\newcommand{\ee}{\end{equation}}
\newcommand{\bea}{\begin{eqnarray}} \newcommand{\eea}{\end{eqnarray}}
\newcommand{\ba}[1]{\begin{array}{#1}} \newcommand{\ea}{\end{array}}

\long\def\symbolfootnote[#1]#2{\begingroup%
\def\thefootnote{\fnsymbol{footnote}}\footnote[#1]{#2}\endgroup} 

\def\eps{\epsilon}

\def\nn{\nonumber}

\newcommand{\cA}{{\cal A}}
\newcommand{\cD}{{\cal D}}
\newcommand{\cM}{{\cal M}}
\newcommand{\cN}{{\cal N}}

\newcommand{\cL}{{\cal L}}

\newcommand{\cK}{{\cal K}}
\newcommand{\cW}{{\cal W}}

\def\bfone{\relax{\rm 1\kern-.35em 1}}

\begin{document}

\begin{titlepage}

\begin{flushright}
\end{flushright}

\bigskip

\begin{center}

\vskip 2cm

{\LARGE \bf The $\textrm{SU(3)}$-invariant sector \\ \vskip .3cm 
of new maximal supergravity} \\

\vskip 1.0cm

{\bf A.~Borghese$^1$, G.~Dibitetto$^2$, A.~Guarino$^3$, D.~Roest$^1$ and O.~Varela$^4$}\\

\vskip 0.5cm

{\em 
$^1$ Centre for Theoretical Physics, University of Groningen, \\ Nijenborgh 4 9747 AG Groningen, The Netherlands \\
$^2$Institutionen f\"or fysik och astronomi, University of Uppsala, \\ Box 803, SE-751 08 Uppsala, Sweden \\
$^3$ Albert Einstein Center for Fundamental Physics, Institute for Theoretical Physics, \\ Bern University, Sidlerstrasse 5, CH–3012 Bern, Switzerland \\
$^4$ Institute for Theoretical Physics and Spinoza Institute, Utrecht University, \\ 3508 TD Utrecht, The Netherlands} \\

\end{center}

\vskip 2cm

\begin{center} {\bf ABSTRACT}\\[3ex]

\begin{minipage}{13cm}
\small

We investigate the $\textrm{SU}(3)$-invariant sector of the one-parameter family of $\textrm{SO}(8)$-gauged maximal supergravities that has been recently discovered. To this end, we construct the $\cN = 2$ truncation of this theory and analyse its full vacuum structure. The number of critical point is doubled and includes new $\cN = 0$ and $\cN =1$ branches. We numerically exhibit the parameter dependence of the location and cosmological constant of all extrema. Moreover, we provide their analytic expressions for cases of special interest. 
Finally, while the mass spectra are found to be parameter independent in most cases, we show that the novel non-supersymmetric branch with $\textrm{SU}(3)$ invariance provides the first counterexample to this.

\end{minipage}

\end{center}

\vspace{2cm}

\vfill

\end{titlepage}

\tableofcontents


\section{Introduction}

Among the gaugings \cite{deWit:2007mt} of maximal supergravity in four dimensions \cite{Cremmer:1979up}, the SO(8) gauging \cite{deWit1} should be singled out as particularly important. The SO(8)-gauged supergravity not only provided the first instance of a complete maximal gauged supergravity, but it also enjoys a clear higher-dimensional origin as a (consistent) truncation of M-theory on the seven-sphere  \cite{deWit2} (see also \cite{Nicolai:2011cy}). Recently, the SO(8)-gauged theory has acquired renewed interest in the light of the AdS$_4$/CFT$_3$ correspondence. The conformal field theory defined on a stack of $N$ M2 branes at an orbifold singularity $\mathbb{C}^4/\mathbb{Z}_k$ has been conjectured in \cite{Aharony:2008ug}, building on \cite{Bagger:2006sk, Gustavsson:2007vu}, to consist in two copies of U$(N) \times$U$(N)$ Chern-Simons theories at levels $k$, $-k$, coupled to bifundamental matter. For $k=1,2$, the ABJM theory has been argued to be maximally supersymmetric, in spite of its superficial $\cN=6$ appearance. Accordingly, for low level, the $\cN=8$ SO(8)-gauged supergravity captures all possible mass terms for the bifundamentals.

On account of the calculational complexity of the full SO(8)-gauged theory, it has proven advantageous to restrict oneself to smaller subsectors invariant under a given subgroup of the full symmetry group. For example, charting the vacuum structure  is usually a much easier task within a smaller subsector than in the full 70-scalar theory\footnote{See nevertheless \cite{Fischbacher:2009cj,Fischbacher:2011jx} for recent progress in the analysis of the vacuum structure of the full  $\cN=8$ SO(8)-gauged potential.}, and complete analytic results can often be found. From a holographic perspective, this (AdS, for the SO(8)-gauging at hand) vacuum structure provides important information about distinct conformal phases of ABJM, while the smaller subsectors themselves map into finite sets of dual field theory operators closed, at least at large $N$, under OPE. These smaller supergravity subsectors are thus extremely valuable to economically assess the dynamics of well-defined finite sets of  field theory operators and, in particular, the field theory's RG evolution upon deformation by relevant operators in this set.

\begin{table}[t] 
\begin{center}
\begin{tabular}{|c|c||r|c|}
\hline 
 SUSY & Symmetry  & Cosm.~constant & Stability\\ 
\hline \hline
$\cN = 8$ & $\textrm{SO}(8)$ & $-6$ ($\times 1$) & $\checkmark$\\[2mm]
\hline
$\cN = 2$ & $\textrm{SU}(3) \times \textrm{U}(1)$ & $-\frac{9}{2}\sqrt{3}$ ($\times 1$) & $\checkmark$\\[2mm]
\hline
$\cN = 1$ & $\textrm{G}_2$ & $-\frac{216}{25}\sqrt{\frac{2}{5}\sqrt{3}}$ ($\times 2$) & $\checkmark$\\[2mm]
\hline
\multirow{2}{*}{$\cN = 0$} & \multirow{2}{*}{$\textrm{SO}(7)$} & $-2\sqrt{5\sqrt{5}}$ ($\times 1$) & $\times$\\[2mm]
 & & $-\frac{25}{8}\sqrt{5}$ ($\times 2$) & $\times$\\[1mm]
\hline
$\cN = 0$ & $\textrm{SU}(4)$ & $-8$  ($\times 1$) & $\times$\\[2mm]
\hline
\end{tabular}
\caption{{\it The $\textrm{SU}(3)$-invariant critical points of the original $\textrm{SO}(8)$-gauged supergravity.}}
\label{Table:SU(3)} 
\end{center}
\vspace{-1cm}
\end{table}

A very interesting subsector of the SO(8)-gauged theory is the $\textrm{SU}(3)$-invariant sector \cite{Warner:1983du, Warner:1983vz}. This is an $\cN=2$ subsector featuring two electric (and two magnetic) vectors and six scalars, organised into one vector and one hypermultiplet. The scalar potential of this theory and its extrema were classified long ago \cite{Warner:1983vz}, although not until recently have some of the corresponding spectra been worked out in the full $\cN=8$ theory \cite{Bobev:2010ib}. The SU(3)-invariant sector features eight AdS critical points (see table \ref{Table:SU(3)}), with various amounts of preserved supersymmetries and with residual gauge symmetry groups including SO(8) down to SU$(3)\times$U(1). In addition, there are several non-supersymmetric points, all of them known to be unstable.\footnote{See  \cite{Fischbacher:2010ec} for an $\textrm{SO}(3) \times \textrm{SO}(3)$-invariant AdS vacuum, thus outside the $\textrm{SU}(3)$ sector, which is non-supersymmetric yet perturbatively stable.} Finally, all these points lift via the consistent embedding \cite{deWit2} to well-known AdS solutions in $D=11$ supergravity, featuring round, squashed, stretched or warped metrics on the internal $S^7$ \cite{Freund:1980xh,Englert:1982vs,deWit:1984va,Pope:1984bd,Corrado}.

The SU(3)-singlet sector of $\cN=8$ SO(8)-gauged supergravity has also proved an extremely fruitful venue for holography. Well before the M2-brane field theory was pinned down, a supersymmetric domain wall interpolating between the $\cN=8$ and $\cN=2$ points was constructed within this sector of the $D=4$ gauged supergravity \cite{Ahn:2000mf} and uplifted to $D=11$ \cite{Corrado}. This was conjectured to holographically describe the RG evolution of the then unknown dual field theory between the corresponding conformal phases. Now, both these UV \cite{Bagger:2006sk,Aharony:2008ug} and IR \cite{Benna:2008zy,Klebanov:2008vq} phases have been determined, and the RG flow between them computed \cite{F-theorem,Benna:2008zy}, with a succesful match between the old \cite{Ahn:2000mf,Corrado} and new \cite{Gabella:2011sg,F-theorem,Gabella:2012rc} supergravity results. Further recent developments in the SU(3)-invariant sector include the construction of various families of RG flows interpolating between all four supersymmetric points \cite{Bobev:2009ms}, and between supersymmetric and non-supersymmetric points \cite{Gauntlett:2009bh}.\footnote{The flows of \cite{Gauntlett:2009bh} were originally constructed in a smaller sector of the $\cN=2$ universal truncation of M-theory on Sasaki-Einstein seven-folds \cite{Gauntlett:2009zw}. It was shown in \cite{Bobev:2010ib} that the theory of \cite{Gauntlett:2009bh} coincides with the SU(4)-invariant sector of SO(8)-gauged supergravity. In other words, M-theory on (skew-whiffed) Sasaki-Einstein \cite{Gauntlett:2009zw} and the SU(3)-invariant sector of SO(8)-gauged supergravity overlap  \cite{Bobev:2010ib} in the theory of \cite{Gauntlett:2009bh} or, equivalently, in the SU(4)-invariant sector of SO(8)-gauged supergravity.} Finally, the SU(3)-invariant sector has proved a very useful arena for top-down AdS/CMT investigations, with holographic superconductivity having been investigated in this model \cite{Donos:2011ut,Bobev:2011rv}.

The SO(8)-gauging of \cite{deWit1}  of $\cN=8$ supergravity \cite{Cremmer:1979up} has always been tacitly assumed to be unique. However, very recently it has been pointed out that, instead, there exists a one-parameter family of $\textrm{SO}(8)$-gauged supergravities \cite{Dall'Agata:2012bb}. All members in the family have the same gauge group, SO(8), with the same embedding into the duality group $\textrm{E}_{7(7)}$, but nevertheless differ from the original theory \cite{deWit1}. The distinguishing feature is the choice of gauge vectors, which can be taken to be electric, magnetic or a dyonic combination thereof, depending on the value of a parameter $\omega$ that can be used to label the theories in the family, with $\omega=0$ the original theory of  \cite{deWit1}. It was shown in \cite{Dall'Agata:2012bb} that all values of $\omega \in [ 0, \pi/8 ]$ lead to inequivalent embedding tensor classifiers, and hence to theories that are not related via $\textrm{E}_{7(7)}$ duality transformations. The much smaller, $\cN=1$ G$_2$-invariant sector of the new theories has been worked out in \cite{Dall'Agata:2012bb} itself, and its vacuum structure determined in \cite{Borghese:2012qm}. Just like its $\omega =0$ counterpart, the $\omega \neq 0$ SU(3)-invariant sector has enormous potential applications which we now set out to explore. In this paper, we will construct the SU(3)-invariant sector of $\omega \neq 0$ SO(8)-gauged supergravity (from its $\omega = 0$ counterpart) and will chart its vacuum structure.  See table \ref{Table:SU(3)2} for a summary.

\begin{table}[t] 
\hspace{-.4cm}
\begin{tabular}{|c|c||r|c||r|c||c|}
\hline 
 SUSY & Symmetry  & CC ($\omega = 0$) & Stability & CC ($\omega = \pi/8$) & Stability & $\omega$-dep. \\ 
\hline \hline
$\cN = 8$ & $\textrm{SO}(8)$ & $-6$ ($\times 1$) & $\checkmark$ & $-6$ ($\times 1$) & $\checkmark$ & $\times$ \\[1mm]
\hline
$\cN = 2$ & $\textrm{SU}(3) \times \textrm{U}(1)$ & $-7.794$ ($\times 1$) & $\checkmark$ & $-8.354$ ($\times 2$) & $\checkmark$ & $\times$ \\[1mm]
\hline
\multirow{2}{*}{$\cN = 1$} & \multirow{2}{*}{$\textrm{G}_2$} & $-7.192$ ($\times 2$) & $\checkmark$ & $-7.943$ ($\times 2$) & $\checkmark$  & $\times$ \\[1mm]
 & & --$\phantom{\textrm{ }(\times 1)}$ & -- & $-7.040$  ($\times 1$) & $\checkmark$ & $\times$ \\[1mm]
\hline
$\cN = 1$ & $\textrm{SU}(3)$ &   --$\phantom{\textrm{ }(\times 1)}$  & -- & $-10.392$ ($\times 1$) & $\checkmark$ & $\times$ \\[1mm]
\hline
\multirow{2}{*}{$\cN = 0$} & \multirow{2}{*}{$\textrm{SO}(7)$} & $-6.687$ ($\times 1$) & $\times$ & $-6.748$ ($\times 2$) & $\times$ & $\times$ \\[1mm]
 & & $-6.988$ ($\times 2$) & $\times$ & $-7.771$ ($\times 2$) & $\times$ & $\times$ \\[1mm]
\hline
$\cN = 0$ & $\textrm{SU}(4)$ & $-8$ ($\times 1$) & $\times$ & $-8.581$ ($\times 2$) & $\times$ & $\times$ \\[1mm]
\hline
$\cN = 0$ & $\textrm{G}_2$ &  --$\phantom{\textrm{ }(\times 1)}$ & -- & $-10.170$ ($\times 1$) &  $\checkmark$ & $\times$ \\[1mm]
\hline
$\cN = 0$ & $\textrm{SU}(3)$ & --$\phantom{\textrm{ }(\times 1)}$ & -- & $-10.237$ ($\times 2$) &  $\checkmark$? & $\checkmark$ \\[1mm]
\hline
\end{tabular}
\caption{{\it The $\textrm{SU}(3)$-invariant critical points of the new $\textrm{SO}(8)$-gauged supergravities. The analytical values of the CC in table~\protect\ref{Table:SU(3)} have been replaced by their approximate numerical values in the $\omega=0$ column 
for comparison's sake. The CC at $\omega = \pi/4$ turns out to coincide with the one at $\omega=0$. An explanation of the question mark concerning the stability of the $\cN=0$ $\textrm{SU}(3)$ points can be found in section~\ref{sec:N=8_spectra}. Finally, the last column indicates the $\omega$-dependence of the mass spectra.}}
\label{Table:SU(3)2} 
\vspace{-.2cm}
\end{table}

We find that the total number of critical points for generic $\omega$ doubles with respect to the $\omega=0$ case. Some of the new points are new 'branches' of the already known $\omega=0$ points, while others are genuinely new for $\omega \neq 0$, and have no counterpart when $\omega =0$. This occurs because the position of the critical points in the $\cN=2$ scalar manifold depends on $\omega$ in such  a way that, when $\omega =0$, some of the critical points are pushed into the boundary of the scalar manifold and thus become unphysical. For example, the unique $\omega=0$ $\cN=2$ point with residual SU$(3)\times$U(1) symmetry partners with a second critical point with the same spectrum, which is unphysical for $\omega=0$. As $\omega$ evolves both points are physical, up until $\omega = \pi/4$, where the situation is reverted: the second point remains physical while the first point exiles to the boundary of moduli space. The evolution in $\omega$ of this pair of critical points is symmetric around $\omega = \pi/8$, a symmetry also observed for all other critical points.

We also find new SU(3)-invariant points,  both $\cN=1$ and non-supersymmetric, with no $\omega =0$ counterpart. Except for the SO(8)-point, not only the location, but also the value of the cosmological constant (CC), namely, the value of the scalar potential at the critical point, varies\footnote{More precisely, the ratio of the CC at each critical point and the CC for the SO(8) point is a function of  $\omega$.} with $\omega$. Finally, it was observed in  \cite{Dall'Agata:2012bb} and then in  \cite{Borghese:2012qm} that the mass spectrum for the critical points with at least G$_2$ invariance is independent of $\omega$. The same conclusion has been reached in \cite{Kodama:2012hu} for a variety of critical points and non-compact gaugings. In the SU(3)-invariant sector of the SO(8)-gauged theory, we find that the spectrum at all points again remains $\omega$-independent, except for the new $\cN=0$ SU(3) points for which we find, for the first time, an $\omega$-dependent spectrum. Moreover, all masses at this point stay above the Breitenlohner-Freedman bound for all $\omega$, both within the SU(3)-invariant sector and, as our preliminary calculations suggest, also in the full $\cN=8$ theory, thus implying stability.

We are in fact able to give results for the spectra in the full $\cN=8$ theory, albeit somewhat loosing control of the actual gauge group. In this context, the embedding tensor formalism turns out to be a very powerful tool for implementing duality covariance\footnote{This formalism was developed in the context of $D=3$ maximal gauged supergravity \cite{Nicolai:2000sc, Nicolai:2001sv}.}. The scan of critical points in such a formalism can be made very systematic by exploiting the homogeneity of the scalar manifold. This feature allows one to restrict the search for solutions to the origin without loss of generality whenever  the considered set of embedding tensor deformations happens to constitute a closed set under non-compact duality transformations. This translates the extremality condition for the scalar potential into a set of quadratic conditions for the deformation parameters. Such a method was first used in ref.~\cite{Dibitetto:2011gm} in the context of $\cN=4$ supergravity and later on in refs~\cite{DallAgata:2011aa, Kodama:2012hu} it was applied for simplicity to some $\cN=8$ cases. Following this approach, the problem of searching for critical points with non-trivial invariance groups can be recast into that of solving a system of quadratic conditions for the set of embedding tensor parameters preserving that symmetry. This was done recently in ref.~\cite{Borghese:2012qm} in the case of $\textrm{G}_{2}$ invariance. Here we extend this analysis to the SU(3)-invariant sector. By comparison with our reduced $\cN=2$ analysis, we are able to pin down the spectra for the SU(3)-invariant points in the entire $\cN=8$ SO(8)-gauged theory. The embedding tensor approach does not commit itself to a specific gauge group though, and thus our scan is able to find Minkowski (but not de Sitter) points corresponding to gaugings other than SO(8).

The organisation of this paper is as follows. We first introduce the general theory for the $\cN =2$ $\textrm{SU}(3)$-invariant sector of maximal supergravity in section \ref{sec:N=2formul}. We demonstrate how the $\omega$-parameter affects the full theory.  Subsequently, in section \ref{sec:VacN=2} we consider the set of supersymmetric and non-supersymmetric critical points within this theory. Moreover, in section \ref{sec:N=8_spectra} we derive the most general supersymmetric $\textrm{SU}(3)$-invariant mass spectra and show that most of these are $\omega$-independent as well. Special attention is paid to the exceptional cases with $\omega$-dependent spectra. In section \ref{sec:FurtherTrunc}, the truncation to the $\cN = 1$ $\textrm{G}_2$ sector is performed. We analyse this truncation and propose a new form of the holomorphic superpotential. Finally, we offer our conclusions and outlook in section \ref{sec:outlook}. We have relegated several technical details to the appendices. Appendix~\ref{N=8_SL8} contains the generalisation of the $\cN = 8$ superpotential to new maximal supergravity. In appendix~\ref{CanonicalN=2} we give the relation between our SU(3)-singlet truncation and the canonical formulation of gauged $\cN=2$ supergravity. In appendix~\ref{app:N=2_vacua} some details about the search of $\cN=2$ vacua in the canonical formalism are provided. Finally, in appendix~\ref{app:other_gaugings}, we give the set of critical points for other gaugings of maximal supergravity.

\medskip

\textbf{Note added:} upon completion of this manuscript we became aware of the preprint \cite{Dall'Agata:2012sx}, which discusses related issues regarding the $\omega$-dependence of solutions for $\textrm{SO}(4,4)$ instead of $\textrm{SO}(8)$ gaugings.

\section{The $\cN=2$ action in electric frame}
\label{sec:N=2formul}

The full $\cN=2$ action, including the vector couplings, of  the $\textrm{SU}(3)$-invariant bosonic sector of the usual $\omega=0$ $\textrm{SO}(8)$-gauged supergravity \cite{deWit1} has been recently given in \cite{Bobev:2010ib}, building on previous partial results. These include the early derivation \cite{Warner:1983vz} of the scalar potential in this sector, and of two superpotentials \cite{Ahn:2000mf} from either of which the potential \cite{Warner:1983vz} derives. The derivation of the action \cite{Bobev:2010ib} strongly relies on the embedding \cite{deWit2} of the full $\cN=8$ theory in $D=11$ supergravity. In the absence of similar explicit embedding formulae for the $\omega \neq 0$ $\textrm{SO}(8)$ gaugings, we will construct the full $\omega$-dependent $\cN=2$ theory
by suitably turning on $\omega$ in the  $\omega=0$ action.

First recall that the $\textrm{SU}(3)$-invariant sector of $\textrm{SO}(8)$-gauged supergravity consists of $\cN=2$ gauged supergravity coupled to a vector and a hypermultiplet. This field content can be obtained by truncating maximal supergravity with respect to a compact $\textrm{SU}(3)$ subgroup of its $\,\textrm{E}_{7(7)}\,$ global symmetry via the chain
\be
\label{chainSU3}
\begin{array}{cccccl}
\textrm{E}_{7(7)} & \supset & \textrm{SL}(2)_{T} \,\times \,\textrm{F}_{4(4)} & \supset & \textrm{SL}(2)_{T} \,\times \,\textrm{SU}(2,1) \,\times \, \textrm{SU}(3) & .
\end{array}
\ee
Such a truncation indeed breaks supersymmetry down to $\cN=2$, as the fundamental representation of $\textrm{SU}(8)$ branches as
\be
\begin{array}{lcll}
\textbf{8} & \rightarrow & \textbf{1}\,\oplus\,\textbf{1}\,\oplus\,\textbf{3}\,\oplus\,\overline{\textbf{3}}  \, ,& 
\end{array}
\ee
thus yielding two invariant gravitini. The $\,\cN=2\,$ truncated theory has four vectors coming from the branching (\ref{chainSU3}) ,
\be
\begin{array}{lcl}
\textbf{56} & \rightarrow & (\textbf{4},\textbf{1},\textbf{1})\,\oplus\,\textrm{non-singlets}  \, ,
\end{array}
\ee
of which only two, the graviphoton and the vector in the vector multiplet, $A^I$, $I=0,1$, are physically independent, the other two, $A_I$, $I=0,1$, being related to them via electromagnetic duality. In the gauged theory, these vectors gauge the $\textrm{U}(1)^2$ that commutes with $\textrm{SU}(3)$ inside $\textrm{SO}(8)$. Finally, the six real scalars of this theory also follow from the decomposition \eqref{chainSU3}: they correspond to the non-compact generators on the right-hand side of that equation, and thus parametrise the following special K\"ahler and quaternionic-K\"ahler manifolds
\be \label{eq:ScalarMan}
\cM_{\textrm{SK}}= \left(\frac{\textrm{SL}(2)}{\textrm{SO}(2)}\right)_{T} \qquad\textrm{and}\qquad \cM_{\textrm{QK}}=\frac{\textrm{SU}(2,1)}{\textrm{SU}(2)_{S} \times \textrm{U}(1)_{U}} \ ,
\ee
associated, respectively, to the vector multiplet and the (universal) hypermultiplet\footnote{We have introduced the labels $_T$, $_{S}$ and $_U$ in (\ref{eq:ScalarMan}) to keep track of the different U(1)'s appearing later on in the text.}. Parametrising the two real scalars in the vector multiplet by a complex coordinate $z$ on the unit disk, and the four real scalars $q^u$, $u=1,\ldots, 4$, in the hypermultiplet by two projective complex coordinates $\left(\zeta_{1},\zeta_{2}\right)$, the metrics on the spaces $\cM_{\textrm{SK}}$ and $\cM_{\textrm{QK}}$ in (\ref{eq:ScalarMan}) read
\begin{eqnarray} \label{ds2VM}
ds^2_{\textrm{SK}} = g_{z \bar z} dz d\bar z \equiv   \frac{3 dz d\bar z }{(1- |z|^2)^2} \, 
\end{eqnarray}
and
\begin{eqnarray} \label{ds2HM}
ds^2_{\textrm{QK}} = h_{uv} dq^ u dq^ v \equiv \frac{d \zeta_1 d\bar{\zeta_1} +  d \zeta_2 d\bar{\zeta_2}}{1-|\zeta_1|^2-|\zeta_2|^2} +  \frac{\big( \zeta_1 d \bar{\zeta_1} + \zeta_2 d \bar{\zeta_2} \big) \big( \bar{\zeta_1} d \zeta_1 + \bar{\zeta_2} d \zeta_2 \big) }{\big(1-|\zeta_1|^2-|\zeta_2|^2\big)^2} \, ,
\end{eqnarray}
respectively. These line elements can be derived \cite{Bobev:2010ib} from the non-linear sigma model of the $\cN=8$ ungauged theory \cite{Cremmer:1979up}. They are thus valid for all gaugings of this $\cN=2$ supergravity model and, in particular, they are $\omega$-independent.

Our starting point to construct the full $\omega$-dependent $\cN=2$ theory is the $\omega=0$ potential. Of the six real scalars in the SU(3)-singlet sector, the scalar potential depends on only four \cite{Warner:1983vz}: those neutral under the gauge group. 
For $\omega =0$ -- and, as will be shown below, also for $\omega \neq 0$ -- the gauging is along a $\textrm{U}(1)_{S} \times \textrm{U}(1)_{U}$ subgroup of the maximal compact subgroup $\textrm{SU}(2)_{S} \times \textrm{U}(1)_{U}$ of the hypermultiplet scalar manifold. The gauge-invariant scalars are thus the special K\"ahler modulus $z$ and a combination of the quaternionic-K\"ahler moduli $q^u=\left(\zeta_{1},\zeta_{2}\right)$ which can be taken to be \cite{Bobev:2010ib} 
\be
\label{zeta12}
\zeta_{12}\,\equiv\,\frac{|\zeta_{1}|+i|\zeta_{2}|}{1+\sqrt{1-|\zeta_{1}|^{2}-|\zeta_{2}|^{2}}} \ .
\ee
The $\omega=0$ theory admits two different superpotentials $\cW_+$ and $\cW_-$ \cite{Ahn:2000mf} which, in the notation of \cite{Bobev:2010ib} (see also \cite{Bobev:2009ms,Ahn:2009as}) read
\begin{align}
\cW_+ = \, & (1 - |z|^{2})^{-3/2} \, (1 - |\zeta_{12}|^{2} )^{-2} \left[ (1 + z^{3}) \, (1 + \zeta_{12}^{4}) + 6 \, z \, (1+z) \, \zeta_{12}^{2} \right] \, ,  \label{SU(3)-Wold}
\end{align}
with $\cW_-$ being obtained from $\cW_+$ by replacing $\zeta_{12}$ with ${\bar \zeta}_{12}$. The potential can thus be written as
\begin{align} \label{PotFromSuperPot}
V =  \,\, & 2 \,\, \bigg[ \frac{4}{3} \, (1 - |z|^{2})^{2} \, \left| \frac{\partial \cW}{\partial z} \right|^{2} + (1 -  |\zeta_{12}|^{2} )^{2} \, \left| \frac{\partial \cW}{\partial \zeta_{12}} \right|^{2} - 3 \cW^{2}  \bigg] \, ,
 \end{align}
where $\cW$ is given by either $|\cW_+|$ or $|\cW_-|$. Note that they are only holomorphic up to the real overall factor. Although both superpotentials $\cW_\pm$ give rise to the same scalar potential, supersymmetric critical points can be extrema of only one or both forms of $\cW$, corresponding to $\cN = 1,2$, respectively.

We find that the scalar potential for the $\omega \neq 0$ theory is still (\ref{PotFromSuperPot}), but with the superpotential now being given, as we will now argue, by either 
\begin{align} 
\label{SU(3) omega superpotential+}
\cW_+ = \, & (1 - |z|^{2})^{-3/2} \, (1 - |\zeta_{12}|^{2} )^{-2} \left[ (e^{2i \omega} + z^{3}) \, (1 + \zeta_{12}^{4}) + 6 \, z \, (1 + e^{2i \omega} z) \, \zeta_{12}^{2} \right] \, , 
\end{align}
or
\begin{align} 
\label{SU(3) omega superpotential-}
\cW_- = \, & (1 - |z|^{2})^{-3/2} \, (1 - |\zeta_{12}|^{2} )^{-2} \left[ (e^{2i \omega} + z^{3}) \, (1 + \bar \zeta_{12}^{4}) + 6 \, z \, (1 + e^{2i \omega} z) \, \bar \zeta_{12}^{2} \right] \, . 
\end{align}
Indeed, the appearance of the phase $\omega$ in the superpotential of the $\textrm{SU}(3)$ sector is unique, up to an overall phase, and can be understood based on the following argument. In \cite{Dall'Agata:2012bb} it was claimed on the basis of an embedding tensor classifier that the new SO(8)-gauged theory is equivalent under a shift of the phase $\omega$ with $\pi/4$, and this was explicitly shown for the $\textrm{G}_2$-invariant sector. In our coordinates, this shift of the phase corresponds to a rotation of $90$ degrees in both the $z$- and the $\zeta_{12}$-plane. In order to realise this symmetry in the $\textrm{SU}(3)$ superpotential, the essentially unique option is to replace \eqref{SU(3)-Wold} with (\ref{SU(3) omega superpotential+}), and similarly for $\cW_-$. Of course, the new superpotential (\ref{SU(3) omega superpotential+})  reduces to  \eqref{SU(3)-Wold} for $\omega = 0$, and passes a number of non-trivial crosschecks. First of all, the resulting potential is compatible with the canonical formulation of $\cN=2$ supergravity (see appendix \ref{CanonicalN=2}). Secondly, it reduces to the  $\textrm{G}_2$-invariant sector potential of \cite{Dall'Agata:2012bb} (see section \ref{sec:FurtherTrunc}). Thirdly, its dependence on the $\textrm{SO}(6)$-invariant dilatons coincides with that of appendix A. Futher crosschecks are listed in the conclusions.


Having pinned down the $\omega$ dependence of the scalar potential, further work is still required to retrieve the rest of the  $\omega$-deformed action from its $\omega=0$ counterpart. Here we just quote the end result, referring to appendix \ref{CanonicalN=2} for the details. The effect of $\omega$ in the full $\cN=8$ theory is to gauge SO(8) dyonically \cite{Dall'Agata:2012bb}. Accordingly, the $\omega=0$ electric frame becomes dyonic for $\omega \neq 0$ and the hyperscalars pick up $\omega$-dependent charges with respect to the electric, $A^I$, and magnetic, $A_I$, vectors of this frame. In this frame, the magnetically charged hyperscalars should appear in the action dualised into two-forms \cite{Louis:2002ny,Theis:2003jj,Dall'Agata:2003yr,D'Auria:2004yi}. Symplectically rotating into a new,  $\omega$-dependent electric frame, thereby eliminating those tensors, we find that the action for the SU(3)-invariant sector of the $\omega$-deformed SO(8)-gauged theory is
\begin{eqnarray} \label{electricN=2action}
{\cal L}& = & \tfrac{1}{2} R *1 + g_{z\bar z} dz \wedge * d\bar z +h_{uv} Dq^u \wedge * Dq^v 
-V*1 \nonumber \\
&& +\tfrac{1}{2} \textrm{Im} \left(\cN_{IJ}^\prime\right)  F^{\prime I} \wedge *F^{\prime J }
+\tfrac{1}{2} \textrm{Re} \left(\cN_{IJ}^\prime\right) F^{\prime I} \wedge F^{\prime J} \, .
\end{eqnarray}
Here, the scalar kinetic terms are governed by the metrics (\ref{ds2VM}), (\ref{ds2HM}), the scalar potential $V$ is obtained from either superpotential (\ref{SU(3) omega superpotential+}) or (\ref{SU(3) omega superpotential-}) via  (\ref{PotFromSuperPot}), and the gauge kinetic matrix $\cN_{IJ}^\prime$ has components
\begin{eqnarray} \label{gaugeKinPrime}
&& \cN_{00}^\prime = i \ \frac{2(z^3+\bar z) + e^{-2i \omega}z^2(3+z\bar z) +  e^{2i\omega}(1+3z\bar z)}{2(z^3-\bar z) + e^{-2i\omega}z^2(3+z\bar z) -  e^{2i\omega}(1+3z\bar z)} \, , \nonumber \\[12pt]
&& \cN_{01}^\prime =\cN_{10}^\prime =  \ \frac{-2i\sqrt{3} z(1+z\bar z) }{2(z^3-\bar z) + e^{-2i\omega}z^2(3+z\bar z) -  e^{2i\omega}(1+3z\bar z)} \, ,  \\[14pt]
&& \cN_{11}^\prime  = i \ \frac{-2(z^3+\bar z) + e^{-2i\omega}z^2(3+z\bar z) +  e^{2i\omega}(1+3z\bar z)}{2(z^3-\bar z) + e^{-2i\omega}z^2(3+z\bar z) -  e^{2i\omega}(1+3z\bar z)} \, . \nonumber
\end{eqnarray}
Finally, in this electric frame, the gauge covariant derivatives of the hyperscalars are
\begin{eqnarray} \label{covDersElectric}
D q^u = d q^u -  A^{\prime I} k^{\prime u}_I \, ,
\end{eqnarray}
where the Killing vectors
\begin{eqnarray} \label{KVsElectric}
k^\prime_0 = i \zeta_1 \partial_{\zeta_1} - i \zeta_2 \partial_{\zeta_2} + \textrm{c.c.} \, , \qquad  
k^\prime_1 = \sqrt{3} i \zeta_1 \partial_{\zeta_1} + \sqrt{3} i \zeta_2 \partial_{\zeta_2} + \textrm{c.c.} \, ,
\end{eqnarray}
generate a compact $\textrm{U}(1)^2$ inside the maximal compact subgroup $\textrm{SU}(2) \times \textrm{U}(1)$ of $\textrm{SU}(2,1)$. 

We have put primes on the electric gauge fields $A^{\prime I}$, $I=0,1$, and their abelian, $F^{\prime I} = dA^{\prime I}$, field strenghts in order to stress that they are expressed in an $\omega$-dependent purely electric frame. They are related to the electric and magnetic gauge fields $(A^I, A_I)$ of the $\omega=0$ electric frame of \cite{Bobev:2010ib} via an $\omega$-dependent Sp$(4,\mathbb{R})$ transformation (see equation (\ref{SympTrans})). For $\omega =0$, $A^{\prime I} =A^I$, and the action (\ref{electricN=2action}) reduces to that of \cite{Bobev:2010ib}. The $\omega=0$ action was shown in that reference to be compatible with the canonical formulation of $\cN=2$ gauged supergravity. In appendix  \ref{CanonicalN=2} we extend this proof to the $\omega \neq 0$ action  (\ref{electricN=2action}).  Furthermore, we discuss the periodicity of this theory in the conclusions.

\section{Vacuum structure and spectra within the $\cN=2$ theory} \label{sec:VacN=2}

Given the particularly simple form of the $\cN = 2$ superpotentials (\ref{SU(3) omega superpotential+}), (\ref{SU(3) omega superpotential-}), it is possible to solve for all its extrema and hence supersymmetric vacua. Barring the maximally supersymmetric $\textrm{SO}(8)$ critical point in the origin $z = \zeta_{12} = 0$, we find essentially three branches of supersymmetric Anti-de Sitter vacua we prepare to describe. In addition, we find four branches of non-supersymmetric AdS vacua. These include all old vacua but also a number of novel branches. Within the $\cN=2$ theory, our critical points are either maximally supersymmetric (in particular, the SO(8) point is only $\cN=2$ within this truncation) or break supersymmetry, partially or totally. See \cite{Hristov:2009uj, Louis:2012ux} for an account of maximally supersymmetric vacua, \cite{Louis:2009xd,Louis:2010ui} for the general conditions for partial supersymmetry breaking, and \cite{Cassani:2012pj} for further recent examples of both cases within the formalism of $\cN=2$ gauged supergravity.\footnote{More generally, see \cite{Meessen:2012sr} for a classification of (time-like) supersymmetric solutions of gauged $\cN=2$ supergravity.}

In addition to the locations of the critical points that we list below, there are additional points related by $\zeta_{12}  \rightarrow - \zeta_{12} $ and $\zeta_{12}  \rightarrow \bar \zeta_{12} $. These points are equivalent and have identical physical properties. This structure of the vacua is a consequence of the even form (in $\zeta_{12} $) of the superpotentials $\cW_\pm$.

\subsubsection*{$\textrm{SU}(3)\times\textrm{U}(1)$-invariant vacua with $\cN = 2$}

\begin{figure*}[ht]
\includegraphics[width=60mm]{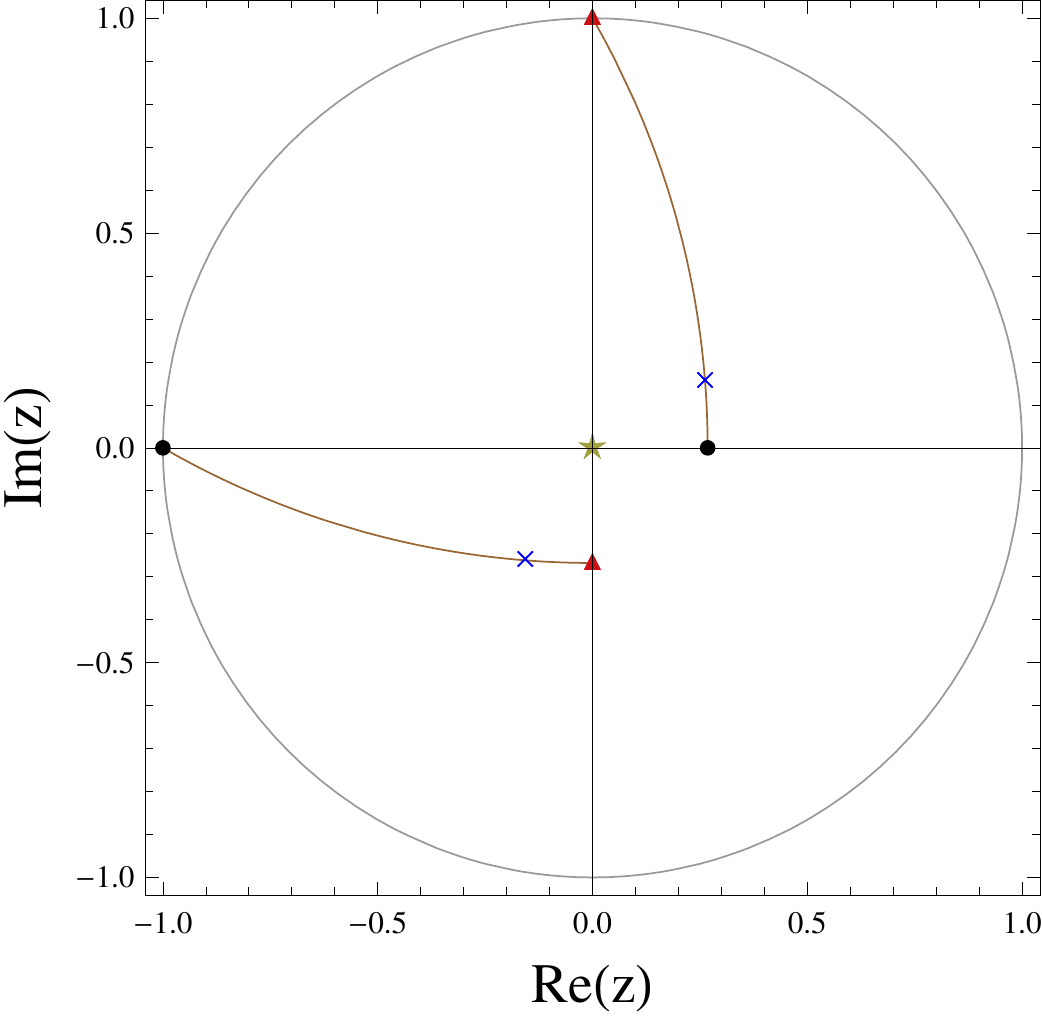}
\hspace{5mm}
\includegraphics[width=88mm]{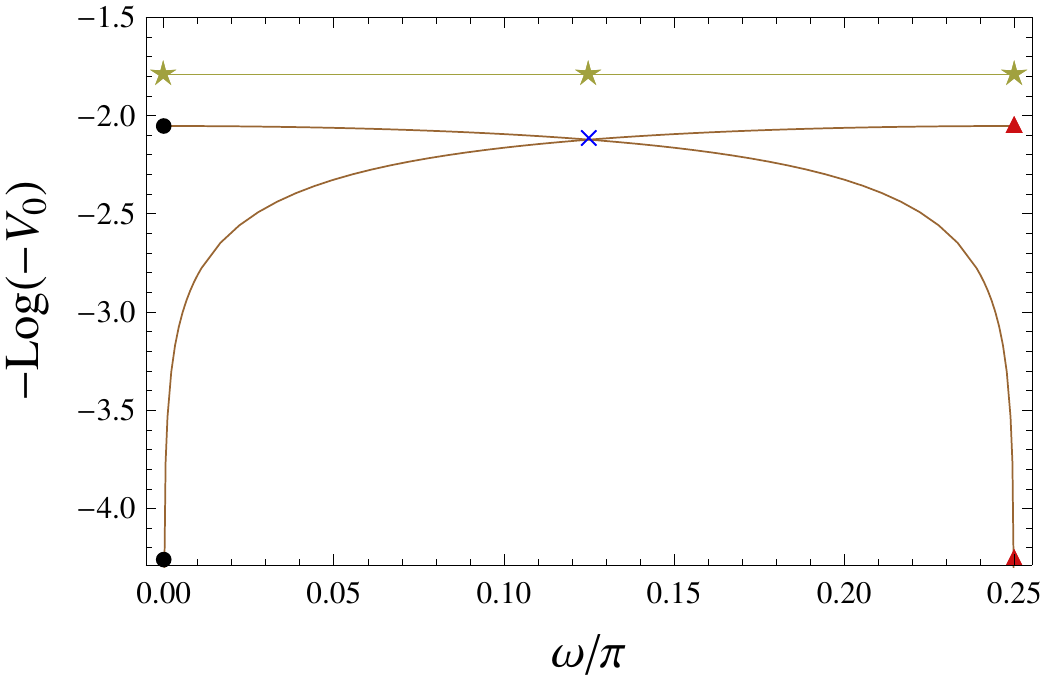}
\vspace{-0.5cm}
\caption{{\it The migration in the $z$-plane (left) and the CC (right) of the $(\textrm{SU}(3) \times \textrm{U}(1))$-invariant solutions preserving $\cN = 2$ as a function of $\omega$: the solutions at $\omega=0$ are denoted by black solid circles, by blue crosses at $\omega=\tfrac18 \pi$ and by red triangles at $\omega=\tfrac14 \pi$.}}
\label{Fig:U(3)_migration} 
\vspace{-0.2cm}
\end{figure*}

\noindent
For generic values of $\omega$, two inequivalent $\cN = 2$ vacua preserving an $\textrm{SU}(3)\times\textrm{U}(1)$ symmetry appear. At the special values of $\omega = n \pi/4$, with $n=0,\pm 1,...$, one of the two solutions becomes singular by migrating to the boundary of the scalar manifold, \textit{i.e.} $|z|=|\zeta_{12}|=1$. The generic behaviour of the two vacua is illustrated in figure~\ref{Fig:U(3)_migration}. It suggests that the vacua structure enjoys a $\pi/4$ periodicity even though this symmetry pattern only holds in figure~\ref{Fig:U(3)_migration} up to an overall $90$ degrees rotation. In view of this mismatch, it is worth mentioning here that the two solutions preserve different $\textrm{U(1)}$ factors in the symmetry group $\textrm{SU}(3)\times\textrm{U}(1)$. 

Here we will present the full analytical expressions for both branches for $\omega$ between $0$ and $\frac{\pi}{4}$. See appendix \ref{app:N=2_vacua} for a derivation of these expressions, and for the analytical formulae for all values of $\omega$. Defining
\begin{eqnarray} \label{defsN=2}
z_0 (\omega) = 1+\frac{4}{(i-\sqrt{3}) |\tan \omega |^{1/3} - 2} \; , \qquad 
f(\omega) = 1+\frac{4}{|\tan \omega |^{2/3} + |\cot \omega |^{2/3} - 1} \,
\end{eqnarray}
and
\begin{eqnarray} \label{eq:PotN=2}
P(\omega) = -3 \sqrt{3} \ \Big( \tfrac{1}{2} | \sin(2\omega) |  +  |\cot \omega |^{1/3} \cos^2 \omega + |\tan \omega |^{1/3} \sin^2 \omega  \Big) \, ,
\end{eqnarray}
the first branch of critical points occurs at
\begin{eqnarray} \label{branch1N=2}
z = -i \ \bar{z}_0 \big(\omega - \tfrac{\pi}{4} \big) \; , \qquad \zeta_{12}^2 =  -|\zeta_{12}|^2 = f\big(\omega - \tfrac{\pi}{4} \big) - \sqrt{ f^2 \big(\omega - \tfrac{\pi}{4} \big) -1 } \, ,
\end{eqnarray}
with cosmological constant $V_0 = P \big(\omega - \tfrac{\pi}{4} \big)$, while the second branch is located at
\begin{eqnarray} \label{branch2N=2}
z =  z_0 (\omega ) \; , \qquad \zeta_{12}^2 =  |\zeta_{12}|^2 = f(\omega) - \sqrt{  f^2( \omega) - 1 } \, ,
\end{eqnarray}
and has cosmological constant $V_0 =P(\omega)$. 

For $\omega = 0$, the branch (\ref{branch2N=2}) lies at infinity and we thus have a unique solution, corresponding to the branch (\ref{branch1N=2}), located at \cite{Bobev:2010ib}
\begin{align}
z = 2-\sqrt{3}  \hspace{8mm},\hspace{8mm}  \zeta_{12} = \pm i \,\left(\sqrt{3} - \sqrt{2}\right) \ ,
\label{WarnerLocation}
\end{align}
and with CC given by
\begin{align}
V_0 = -\frac{9 \sqrt{3}}2{} \, .
\label{CCWarner}
\end{align}
For intermediate values of $\omega$ between $0$ and $\frac{\pi}{4}$ both branches (\ref{branch1N=2}) and (\ref{branch2N=2}) are physical and related through
\begin{equation}
z \rightarrow -i \, \bar{z} 
\hspace{10mm} \textrm{and} \hspace{10mm} 
\zeta_{12} \rightarrow i \, \bar{\zeta}_{12} \ .
\label{transf_sym}
\end{equation}
For $\omega = \frac{\pi}{8}$, the two distinct critical points have equal CC, given by
\begin{align} 
V_0 = -\tfrac{3 \sqrt{6}}{4} \left(  1 + \big( \sqrt{2} +1 \big)^{4/3} + \big( \sqrt{2} -1 \big)^{4/3} \right) \, ,
\label{CCN=2}
\end{align}
and correspond to the crossing of branches in figure~\ref{Fig:U(3)_migration} (right). Finally, when $\omega = \pi/4$, the branch (\ref{branch1N=2}) disappears from the physical scalar manifold and again a single solution remains, corresponding to the branch (\ref{branch2N=2}), located at
\begin{align}
z = -i (2-\sqrt{3})  \hspace{8mm},\hspace{8mm}  \zeta_{12} = \pm (\sqrt{3} - \sqrt{2}) \ ,
\end{align}
and with the same CC as in (\ref{CCWarner}).

The scalar masses and conformal dimensions in the $\cN = 2$ sector are given by\footnote{Throughout the paper, all masses of scalars and vectors in AdS critical points will be normalised w.r.t. the AdS radius $L^2=-3/V_{0}$.}
\be
\begin{array}{cccccccc}
m^2 L^2 & =  & 3 \pm \sqrt{17} \quad (\times 1) & , &  2 \quad (\times 3) & , &  0  \quad (\times 1) & , \\
\Delta & = & \frac{1}{2} \left( 3 \pm 2 + \sqrt{17} \right) & , &  \frac{1}{2} \left( 3 + \sqrt{17} \right) & , & \text{unphysical} & , 
\end{array} 
\ee
while the vector masses  and dimensions are

\be
\begin{array}{cccccc}
m^2 L^2 & = &  4 \quad (\times 1) & , & 0 \quad (\times 1) & , \\
\Delta & = & \frac{1}{2} \left( 3 + \sqrt{17} \right) & , & 2 & . 
\end{array}
\ee
The full spectra can be found in section 4. Note in particular that these are independent of $\omega$. This spectrum fits in $\textrm{OSp}(2|4)$ multiplets. First of all the massless scalar is eaten up by one of the vectors which becomes massive, thereby breaking the $\textrm{U}(1) \times \textrm{U}(1)$ gauge symmetry down to a single $\textrm{U}(1)$. The massless vector sits in the $\cN = 2$ supergravity multiplet while the massive vector together with the remaining five scalars fill out a long vector multiplet.

\subsubsection*{$\textrm{G}_2$-invariant vacua with $\cN = 1$}

\begin{figure*}[ht]
\begin{center}
\includegraphics[width=60mm]{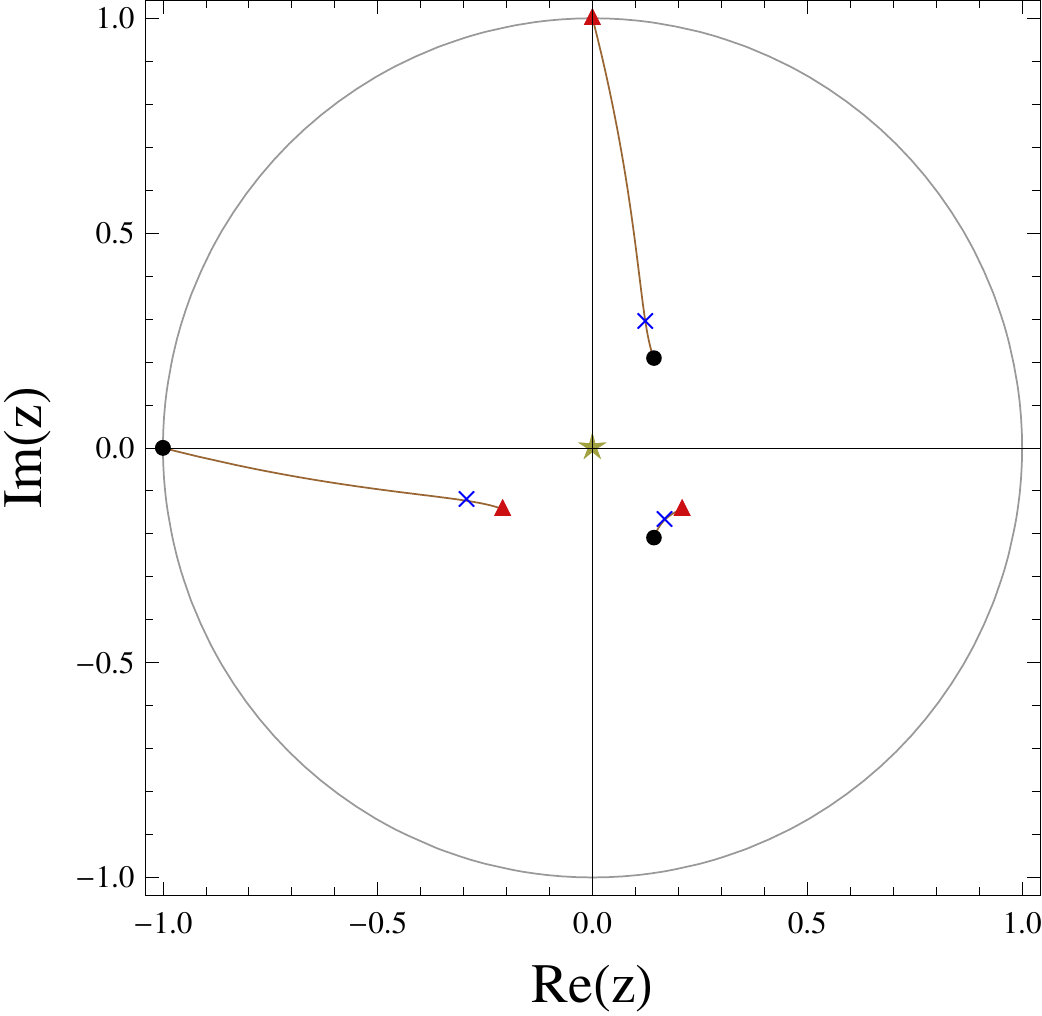}
\hspace{5mm}
\includegraphics[width=84mm]{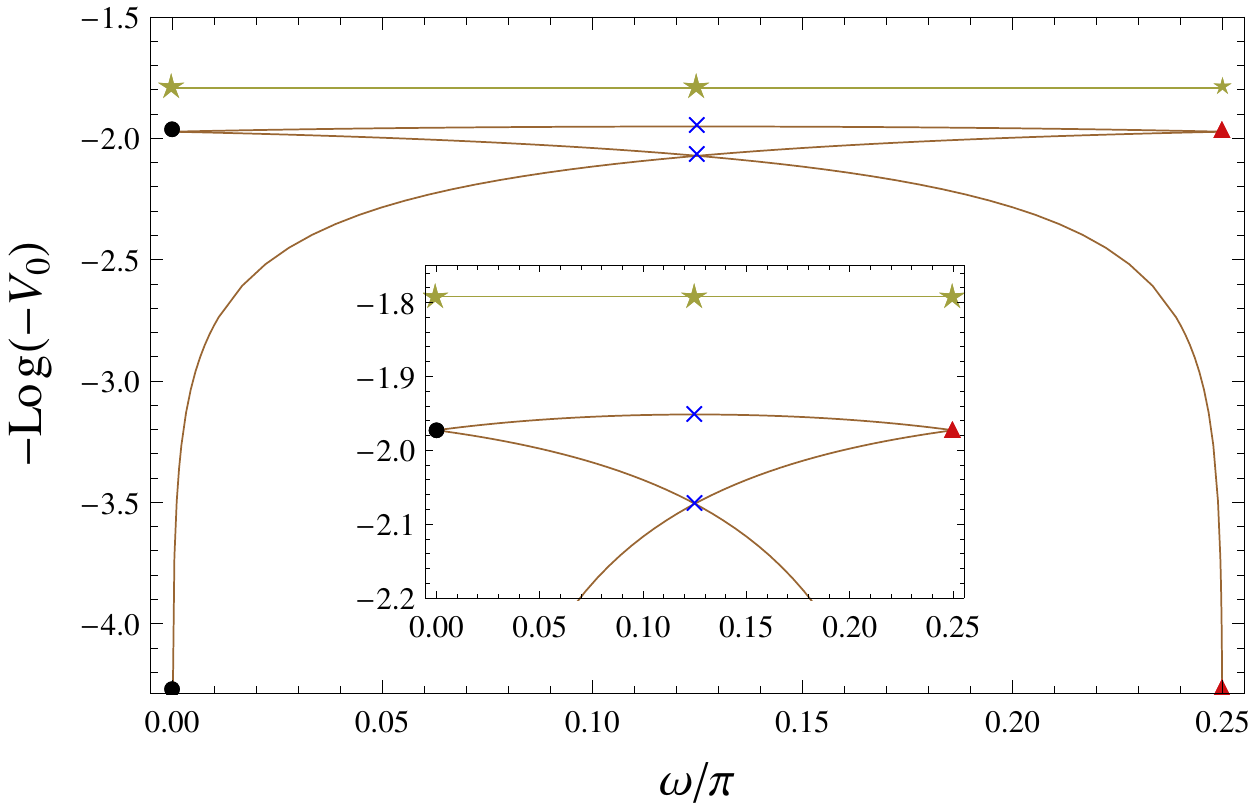} 
\vspace{-0.5cm}
\caption{{\it  The migration in the $z$-plane (left) and the CC (right) of the three $\textrm{G}_{2}$-invariant solutions preserving $\cN = 1$ as a function of $\omega$: the solutions at $\omega=0$ are denoted by black solid circles, by blue crosses at $\omega=\tfrac18 \pi$ and by red triangles at $\omega=\tfrac14 \pi$.}}
\label{Fig:G2_migration}
\end{center}
\vspace{-0.5cm}
\end{figure*}

\noindent
The $\textrm{G}_{2}$ truncation is compatible with the identification $\zeta_{12}=z$. In this case, there are three different $\cN = 1$ vacua preserving a $\textrm{G}_2$ symmetry for generic values of $\omega$. One of these migrates to the boundary when $\omega=n \pi/4$. Moreover, two of the points are parity symmetric: $i)$ $z \rightarrow \bar z$ at $\omega = 0$, $\,\,ii)$ there is a diagonal symmetry for $\omega = \pi/8$, $\,\,iii)$ $z \rightarrow -\bar z$ at $\omega = \pi/4$, in agreement with ref.~\cite{Dall'Agata:2012bb}. The generic behaviour of these $\textrm{G}_2$-invariant solutions is illustrated in figure~\ref{Fig:G2_migration}.

For the standard choice of $\omega = 0$ in the superpotential, the most general solution is given by a pair of $\mathbb{Z}_2$-related points \cite{Bobev:2010ib}
\begin{equation}
\begin{array}{ccl}
z \,=\, \zeta_{12} & = & \dfrac{1}{4} \left( 1 \pm \dfrac{i}{3^{1/4}} \sqrt{2 + \sqrt{3}} \right) (3 + \sqrt{3} - 3^{1/4} \sqrt{10}) \ ,
\end{array}
\label{G2Location}
\end{equation}
with energy $V_{0}=-\frac{216}{25}\sqrt{\frac{2}{5}\sqrt{3}}$. For the very special value of $\omega = \pi/8$ we find the three inequivalent points
\begin{equation}
z = \zeta_{12} =  \tfrac{1}{2} \left(\sqrt{2}+\sqrt{3}-\sqrt{3+2 \sqrt{6}}\right) (1-i) 
\end{equation}
and
\begin{equation}
z = \zeta_{12} =  0.123 + i \, 0.293  
\hspace{10mm} \textrm{,} \hspace{10mm}
z = \zeta_{12} =  -0.293 - i \, 0.123 \ ,
\end{equation}
the latter being related by (\ref{transf_sym}). Finally, for $\omega=\pi/4$, one finds a pair of $\mathbb{Z}_2$-related points. These are related to (\ref{G2Location}) again by applying (\ref{transf_sym}) and produce the same value of the cosmological constant, as shown in figure~\ref{Fig:G2_migration}.

For any value of the phase $\omega$, the masses of the scalar fields and the conformal dimensions are given by
\be
\begin{array}{cccccccc} \label{eq:ConDimsG2}
m^2 L^2  & = & 4 \pm \sqrt{6} \quad (\times 1) & , & - \frac{1}{6} \left( 11 \pm \sqrt{6} \right) \quad (\times 1) & , &  0 \, \quad (\times 2) & , \\
\Delta & = & \frac{1}{2} \left( 3 \pm 1 +2 \sqrt{6} \right) & , & 2 \mp \frac{1}{\sqrt{6}} & , & \text{unphysical} & ,
\end{array}
\ee
for all the critical points above. Similarly, the vector fields have a mass
\be
m^2 L^2 = 
\begin{array}{cc}
\frac{1}{2} \left(3 \pm \sqrt{6} \right) \quad (\times 1) & .
\end{array}
\ee
In this case these span the following $\textrm{OSp}(1|4)$ multiplets: two massless scalars are eaten up by the two vectors. Being supersymmetry broken down to $\cN = 1$, one of the gravitini becomes massive and form a massive gravitino multiplet together with the two vectors and with one spin-1/2 field. The scalars pair up two by two in chiral multiplets. Note that, although we have only quoted the largest root $\Delta$ of $m^2 L^2 = \Delta (\Delta -3)$ in (\ref{eq:ConDimsG2}), the pairing into a chiral multiplet for the fields with mass $m^2 L^2  = - \frac{1}{6} \left( 11 \pm \sqrt{6} \right)$ requires also the shorter root.

\subsubsection*{$\textrm{SU}(3)$-invariant vacua with $\cN = 1$}

\noindent
For generic values of $\omega$, there is one $\cN = 1$ point preserving an $\textrm{SU}(3)$ symmetry which migrates to the boundary for $\omega=n \pi/4$. This is a solution of the $\textrm{SO}(8)$ gauging which could not be found before since it disappears for the standard choice of $\omega=0$. Thus, it is a genuine supersymmetric solution of the new maximal supergravity\footnote{The existence of genuinely new G$_{2}$-preserving solutions of new maximal supergravity was originally noticed in ref.~\cite{Dall'Agata:2012bb} and further confirmed by the computation of the vector and scalar mass spectra in ref.~\cite{Borghese:2012qm}, where they were also found to be non-supersymmetric.}.

\begin{figure*}[ht]
\includegraphics[width=60mm]{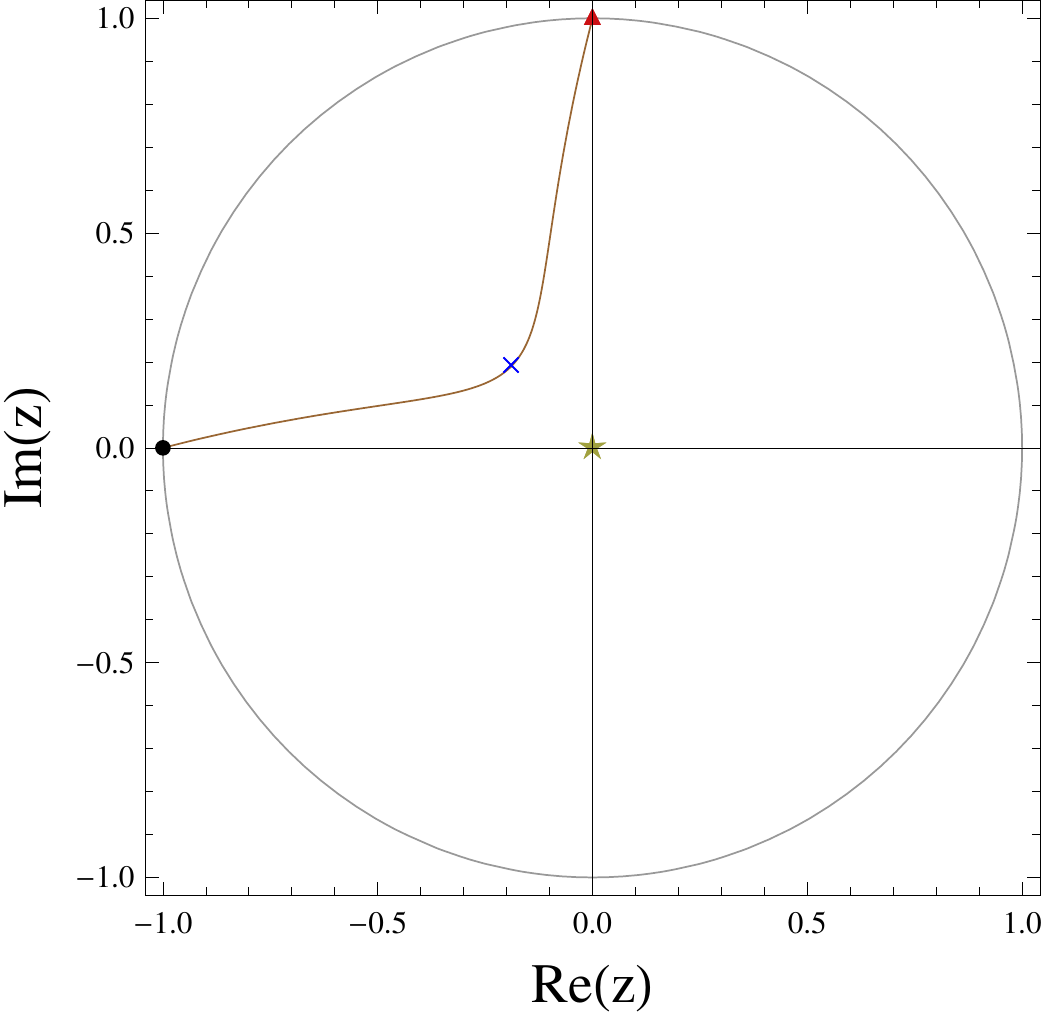}
\hspace{5mm}
\includegraphics[width=88mm]{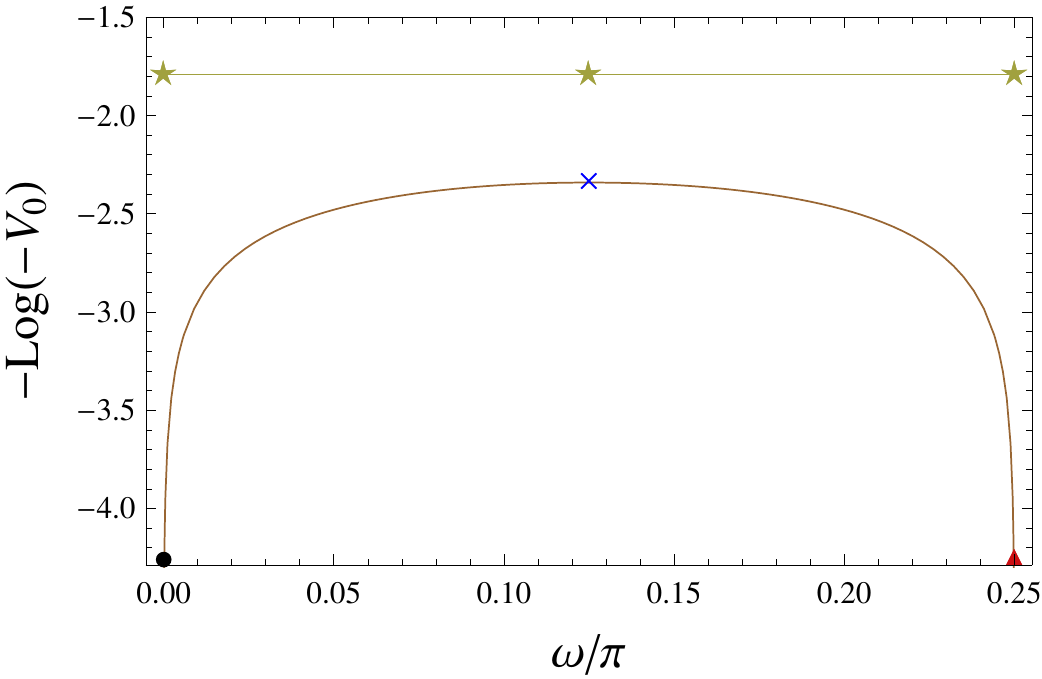} 
\vspace{-0.5cm}
\caption{{\it  The migration in the $z$-plane (left) and the CC (right) of the $\textrm{SU}(3)$-invariant point with $\cN = 1$ as a function of $\omega$: the solutions at $\omega=0$ are denoted by black solid circles, by blue crosses at $\omega=\tfrac18 \pi$ and by red triangles at $\omega=\tfrac14 \pi$.}}
\label{Fig:SU(3)_migration} 
\vspace{-0.5cm}
\end{figure*}

For $\omega = 0$, there are no acceptable solutions as the fields lie at the boundary of the scalar manifold. As long as $\omega$ is turned on, one solution appears which flows towards the location
\begin{align}
z = \left( \sqrt{\tfrac{3}{2}}-\sqrt{2} \right) (1 - i)  \hspace{6mm},\hspace{6mm}  \zeta_{12} = \frac{1}{2} \left(1-\sqrt{3}\right) ( 1+ i) \ ,
 \end{align}
with energy $V_{0}=-6 \sqrt{3}$ at the very special value of $\omega=\pi/8$. Finally, when ${\omega=\pi/4}$, the fields move again to the boundary of the scalar manifold and the solution disappears. This behaviour is illustrated in figure~\ref{Fig:SU(3)_migration}.

The scalar mass spectrum and conformal dimensions for this $\cN = 1$ sector are given by
\be
\begin{array}{cccccc}
m^2 L^2 & = & 4 \pm \sqrt{6} \quad (\times 2) & , & 0 \quad (\times 2) & , \\
\Delta  & = & \frac{1}{2} \left( 3 \pm 1 +2 \sqrt{6} \right) & , & \text{unphysical} & , 
\end{array}
\ee
while the vector masses are
\be
m^2 L^2 =
\begin{array}{cccc}
 2 \quad (\times 1) & , &  6 \quad (\times 1) & .
\end{array}
\ee
The full spectra are given also in section 4. The two vectors belong to a $\textrm{OSp}(1|4)$ massive gravitino multiplet while the non zero scalars belong to two chiral multiplets. The massless scalars are eaten up by the vectors. Once more, all the masses happen to be independent of $\omega$.

\subsubsection*{$\textrm{SO}(7)$-invariant non-supersymmetric vacua}

\begin{figure*}[t]
\begin{center}
\includegraphics[width=60mm]{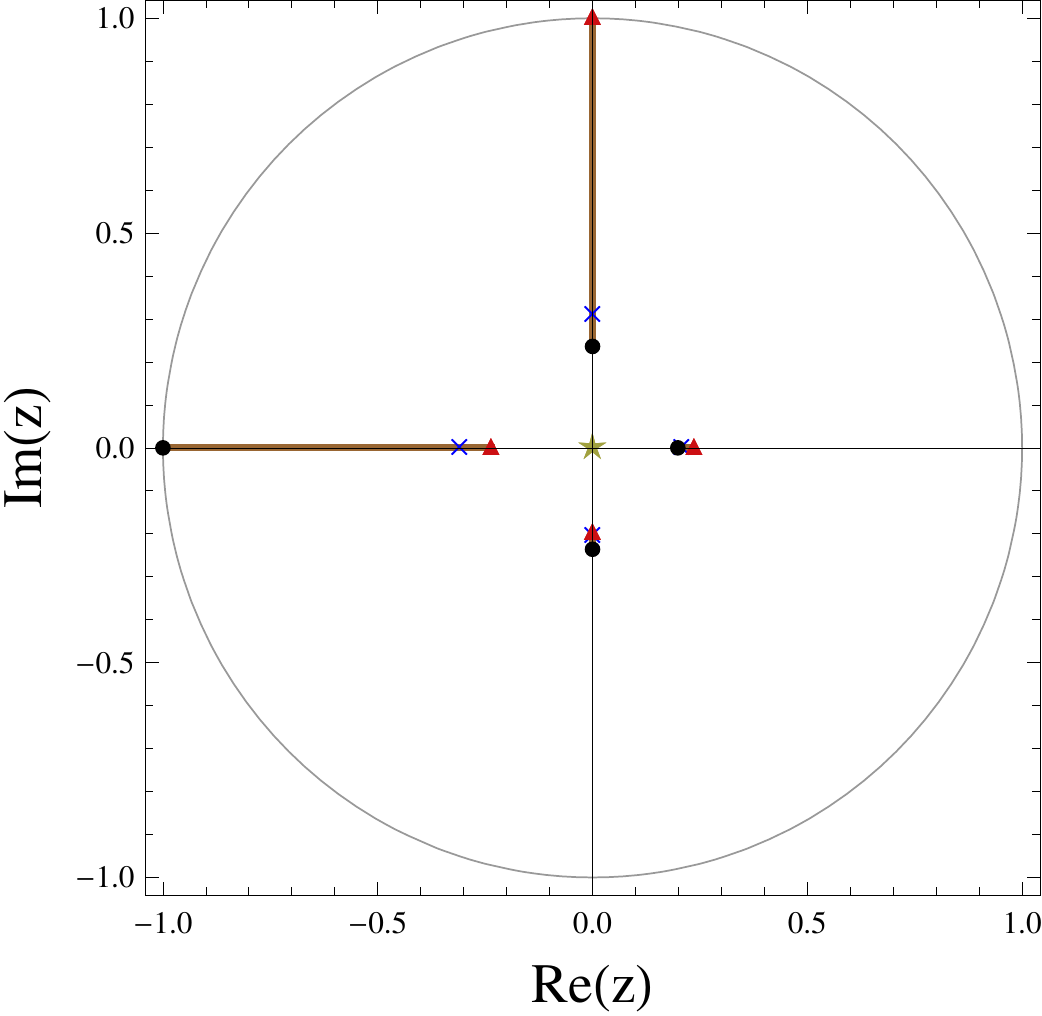}
\hspace{5mm}
\includegraphics[width=86mm]{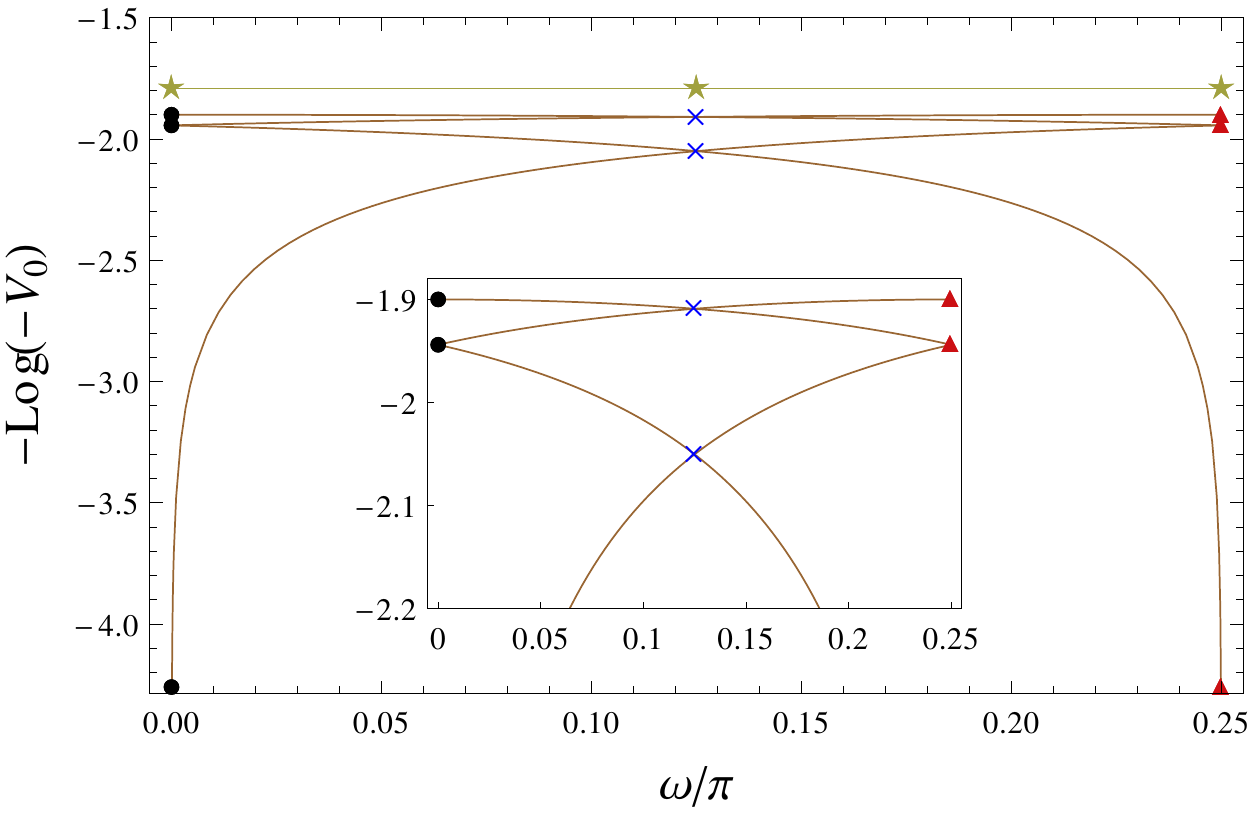} 
\vspace{-0.5cm}
\caption{{\it  The migration in the $z$-plane (left) and the CC (right) of the four $\textrm{SO}(7)$-invariant solutions preserving $\cN = 0$ as a function of $\omega$: the solutions at $\omega=0$ are denoted by black solid circles, by blue crosses at $\omega=\tfrac18 \pi$ and by red triangles at $\omega=\tfrac14 \pi$.}}
\label{Fig:SO(7)_migration}
\end{center}
\vspace{-1cm}
\end{figure*}

For a generic value of $\omega$ there are four inequivalent $\textrm{SO}(7)$-invariant and non-supersymmetric critical points. As an example, $\omega=\pi/8$ produces critical points located at
\be
\begin{array}{lclc}
z=\zeta_{12}=0.207     & \hspace{6mm} , \hspace{6mm} &  z=-\zeta_{12}=- i\, 0.207 & , \\[2mm]
z=\zeta_{12}=i\, 0.310   & , &  z=-\zeta_{12}=-0.310 & ,
\end{array}
\ee
with energies $V_{0}=-6.748$ (upper line) and $V_{0}=-7.771$ (lower line). The two points in the upper (equivalently lower) line are connected via the transformation in (\ref{transf_sym}). At the special values of $\omega = n \pi / 4$, one of the four points migrates to the boundary of the scalar manifold and another two become degenerate in energy. For instance, $\omega=0$ gives rise to critical points at
\be
\label{SO7_w=0}
z=\zeta_{12}=0.199   
\hspace{10mm} \textrm{and} \hspace{10mm}
z=\zeta_{12}=\pm i\, 0.236
\ee
with energies $V_{0}=-6.687$ and $V_{0}=-6.988$, respectively. In a similar way, $\omega=\pi/4$ does it at
\be
\label{SO7_w=pi/4}
z=- \zeta_{12}=\mp 0.236   
\hspace{10mm} \textrm{and} \hspace{10mm}
z=-\zeta_{12}=- i\, 0.199
\ee
with energies $V_{0}=-6.988$ and $V_{0}=-6.687$, respectively. The $\omega$-evolutions of critical point positions and CC are shown in figure~\ref{Fig:SO(7)_migration}. Notice that the solutions (\ref{SO7_w=0}) and (\ref{SO7_w=pi/4}) are again related by the field transformations in (\ref{transf_sym}). These solutions belong also to the $\textrm{G}_2$-invariant truncation of the theory and hence belong to the truncation first presented in ref.~\cite{Dall'Agata:2012bb}. The full $\cN=8$ spectrum is the same for all of them and was found to be $\omega$-independent in ref.~\cite{Borghese:2012qm}. For the sake of completeness, we have included it in the next section.

Within the $\textrm{SU}(3)$ truncation, the scalar spectrum consists of
\be
\begin{array}{cccccccccc}
m^2 L^2 & = & 6 \quad (\times 1) & , & -\frac{12}{5} \quad (\times 1) & , & - \frac{6}{5} \quad (\times 3) & , & 0 \quad (\times 1) & ,
\end{array}
\ee
while the vector masses are
\be
m^2 L^2 = 
\begin{array}{cccc}
\dfrac{12}{5} \quad (\times 1) & , & 0 \quad (\times 1) & .
\end{array}
\ee

\subsubsection*{$\textrm{G}_2$-invariant non-supersymmetric vacuum}

\begin{figure*}[h!]
\begin{center}
\includegraphics[width=60mm]{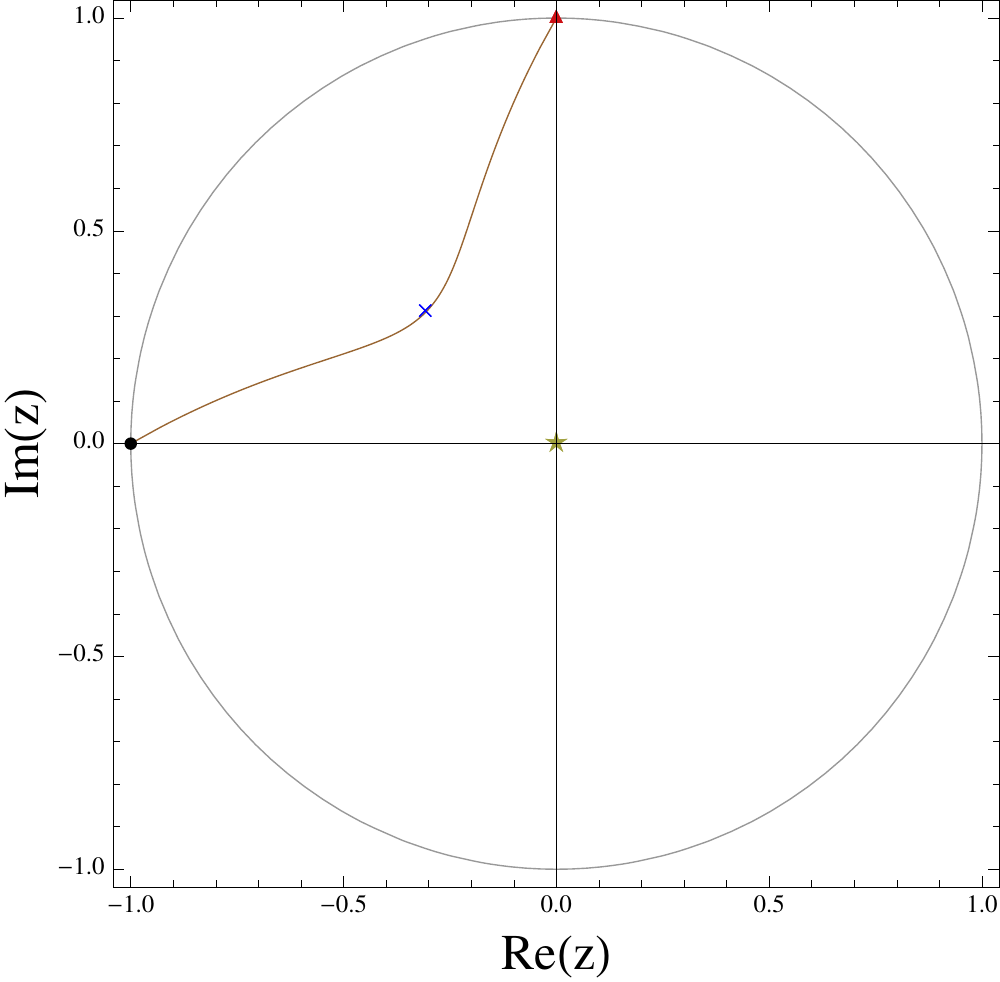}
\hspace{5mm}
\includegraphics[width=86mm]{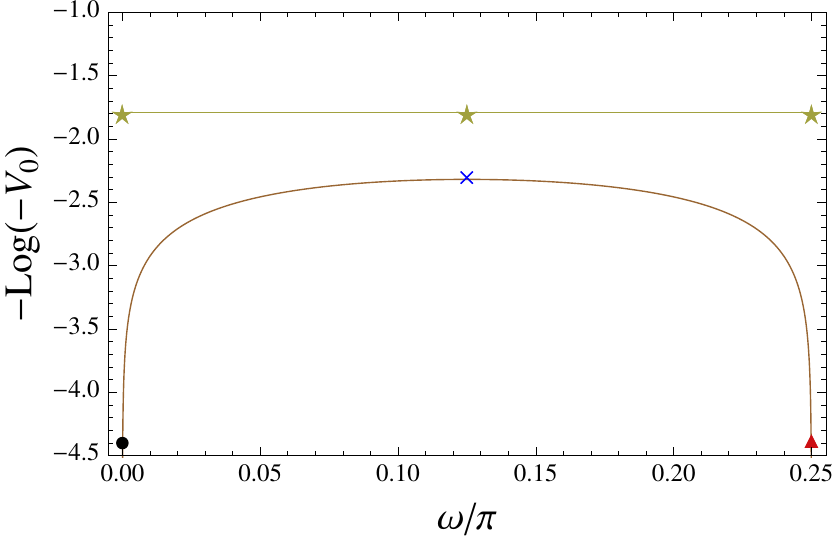} 
\vspace{-0.5cm}
\caption{{\it  The migration in the $z$-plane (left) and the CC (right) of the $\textrm{G}_{2}$-invariant solution preserving $\cN = 0$ as a function of $\omega$: the solutions at $\omega=0$ are denoted by black solid circles, by blue crosses at $\omega=\tfrac18 \pi$ and by red triangles at $\omega=\tfrac14 \pi$.}}
\label{Fig:G2nonSUSY_migration}
\end{center}
\vspace{-0.5cm}
\end{figure*}

For a generic value of $\omega$, there is one inequivalent $\textrm{G}_2$-invariant and non-supersymmetric critical point. For instance, when $\omega=\pi/8$, the critical point is located at
\be
z=\zeta_{12}=-0.308 + i\, 0.308 \ ,
\ee
and has a vacuum energy $V_{0}=-10.170$. This point migrates to the boundary of the scalar manifold at the special values of $\omega = n \pi / 4$, as shown in figure~\ref{Fig:G2nonSUSY_migration}. Therefore, this solution represents a genuine new maximal supergravity solution. It belongs to the $\textrm{G}_2$-invariant truncation too and hence was originally noticed in ref.~\cite{Dall'Agata:2012bb}. The full $\cN=8$ spectrum happens to be $\omega$-independent and was first computed in ref.~\cite{Borghese:2012qm}. We have also included it in the next section.

When restricted to the $\textrm{SU}(3)$ sector, the scalar spectrum is given by
\be
\begin{array}{cccccccc}
m^2 L^2 & = & 6 \quad (\times 2) & , & -1 \quad (\times 2) & , & 0 \quad (\times 2) & , 
\end{array}
\ee
and the vector masses read
\be
m^2 L^2 = 
\begin{array}{cc}
3 \quad (\times 2) & .
\end{array}
\ee

\subsubsection*{$\textrm{SU}(4)$-invariant non-supersymmetric vacua}

\begin{figure*}[t]
\begin{center}
\includegraphics[width=60mm]{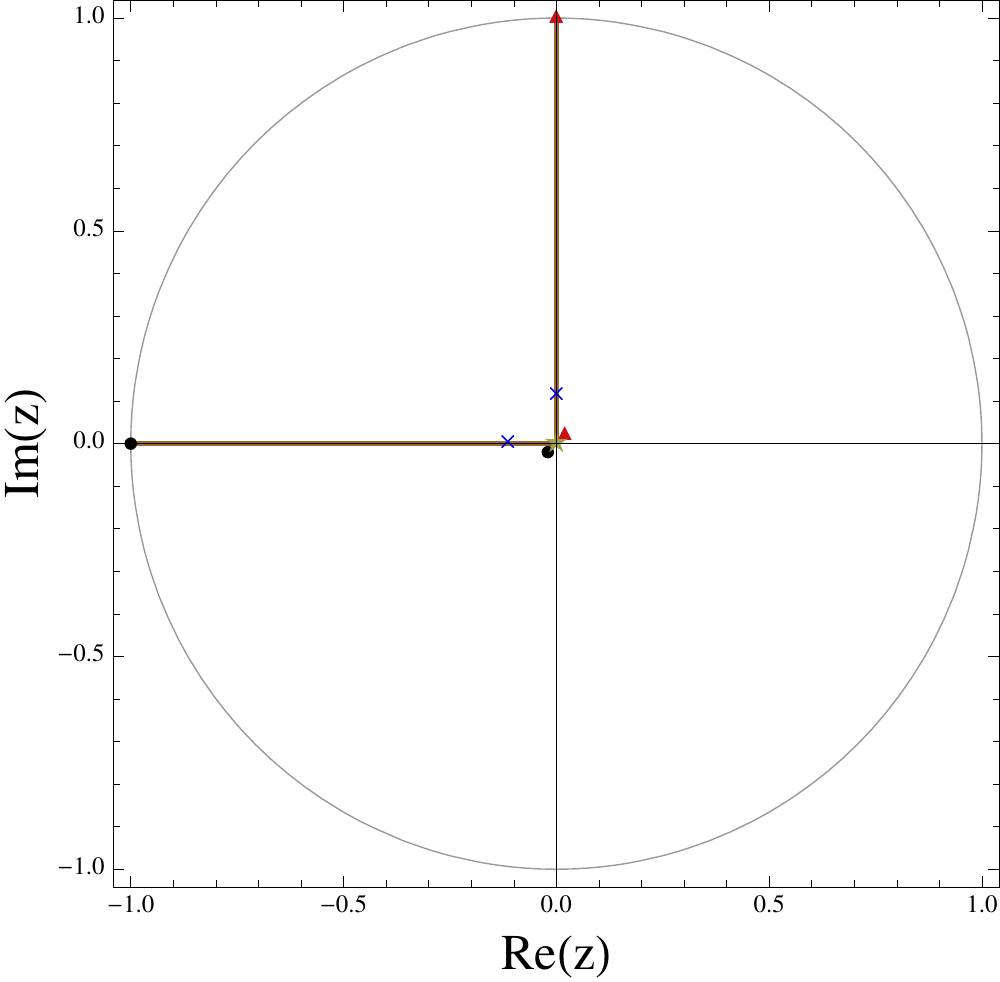}
\hspace{5mm}
\includegraphics[width=86mm]{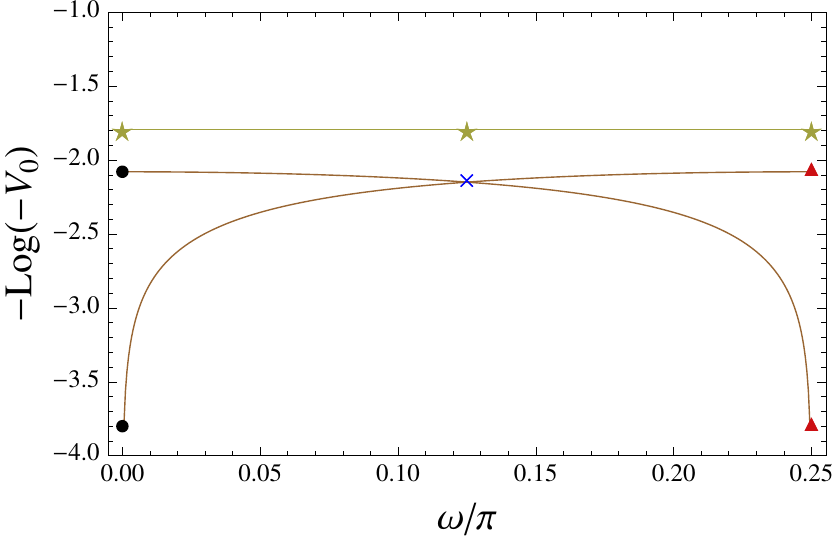} 
\vspace{-0.5cm}
\caption{{\it  The migration in the $z$-plane (left) and the CC (right) of the two $\textrm{SU}(4)$-invariant solutions preserving $\cN = 0$ as a function of $\omega$: the solutions at $\omega=0$ are denoted by black solid circles, by blue crosses at $\omega=\tfrac18 \pi$ and by red triangles at $\omega=\tfrac14 \pi$.}}
\label{Fig:SU4nonSUSY_migration}
\end{center}
\vspace{-0.5cm}
\end{figure*}

There are two $\textrm{SU}(4)$-invariant and non-supersymmetric critical points for generic values of the $\omega$ parameter. As long as $\omega$ changes, the location of these critical points also varies. For $\omega=\pi/8$, the two solutions are located at
\be
z=-0.114  \,\,\,,\,\,\, \zeta_{12}=0.453
\hspace{10mm} \textrm{and} \hspace{10mm}
z=i\, 0.114 \,\,\,,\,\,\, \zeta_{12}=i\, 0.453 \ .
\ee
They are again related by (\ref{transf_sym}) and become degenerate in energy with $V_{0}=-8.581$. Setting $\omega = 0$, one solution was first discovered in ref.~\cite{Warner:1983vz}
\be
z=0 \,\,\,\,\,\,\,,\,\,\,\,\,\,\, \zeta_{12}=(\sqrt{2}-1) i \ ,
\ee
with $V_{0}=-8$, whereas the other becomes singular by moving to the boundary of the moduli space. At the critical value $\omega = \pi/4$, the situation at $\omega = 0$ is recovered but with the role of the two points exchanged
\be
z=0 \,\,\,\,\,\,\,,\,\,\,\,\,\,\, \zeta_{12}=(\sqrt{2}-1) \ .
\ee
Once more, the solutions at $\omega=0$ and $\omega=\pi/4$ are related via the field transformations in (\ref{transf_sym}) and their behaviour is shown in figure~\ref{Fig:SU4nonSUSY_migration}.

The scalar masses within the $\textrm{SU(3)}$ truncation are
\be
\begin{array}{cccccccc}
m^2 L^2 & = & 6 \quad (\times 2) & , & -\frac{3}{4} \quad (\times 2) & , & 0 \quad (\times 2) &  , 
\end{array}
\ee
whereas those of the vectors read
\be
m^2 L^2 = 
\begin{array}{cccc}
6 \quad (\times 1) & , & 0 \quad (\times 1) & ,
\end{array}
\ee
hence being $\omega$-independent as well.

\subsubsection*{$\textrm{SU}(3)$-invariant non-supersymmetric vacua}

\begin{figure*}[t]
\includegraphics[width=60mm]{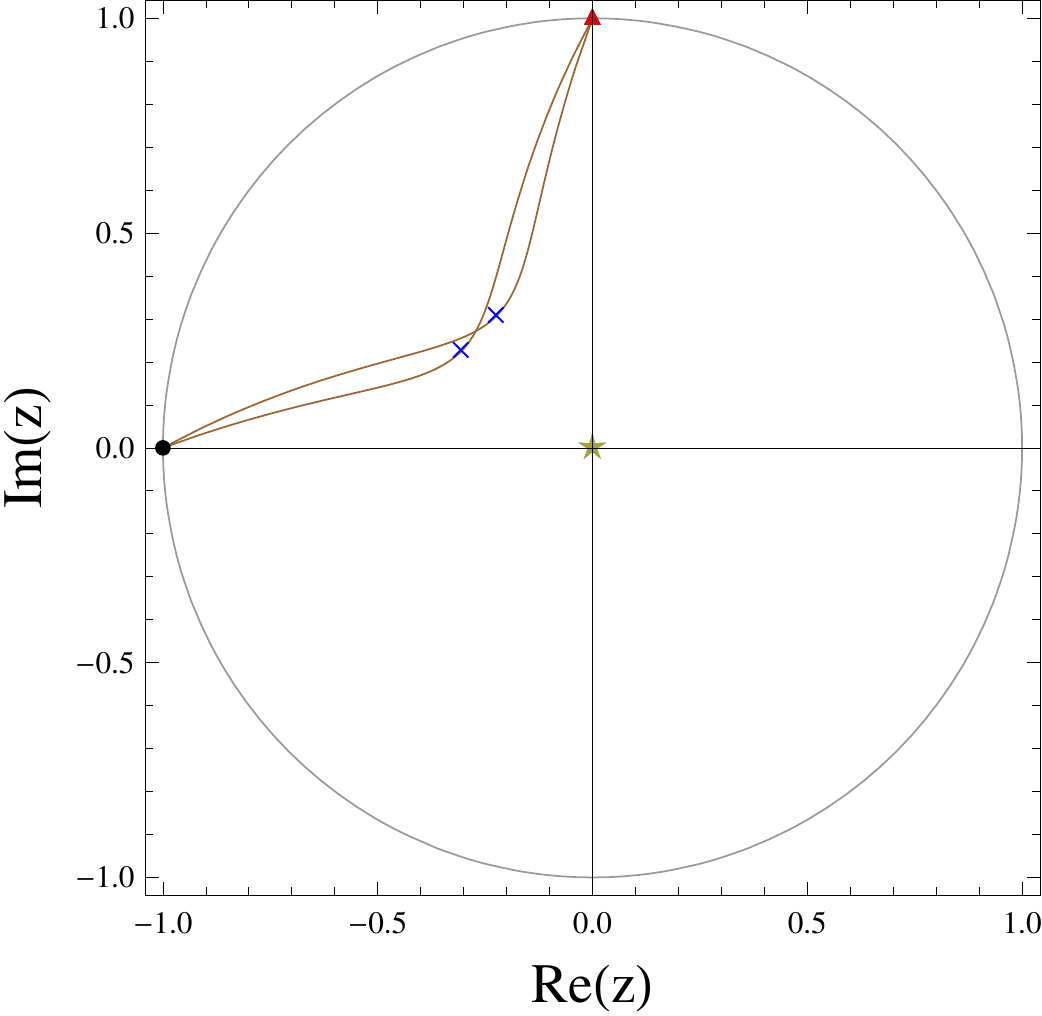}
\hspace{5mm}
\includegraphics[width=90mm]{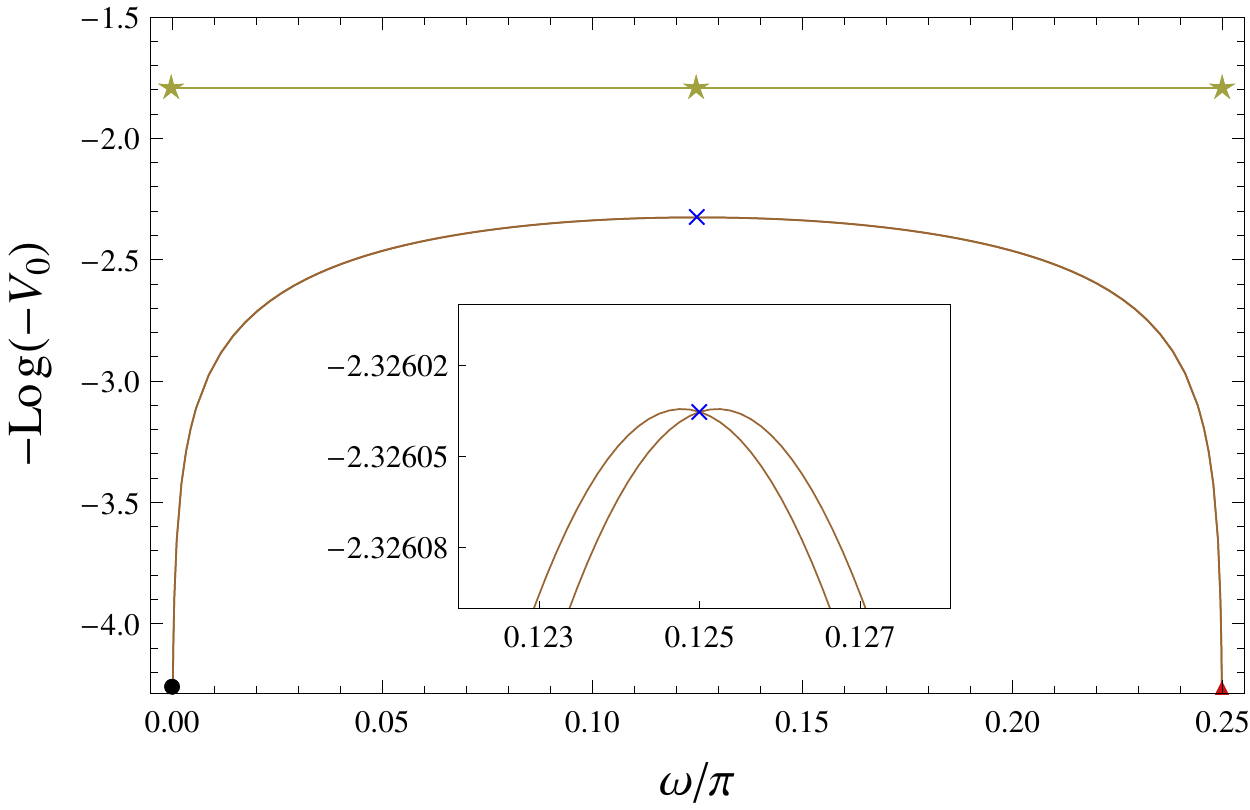}
\vspace{-0.5cm}
\caption{{\it  The migration in the $z$-plane (left) and the CC (right) of the two non-supersymmetric $\textrm{SU}(3)$-invariant points as a function of $\omega$: the solutions at $\omega=0$ are denoted by black solid circles, by blue crosses at $\omega=\tfrac18 \pi$ and by red triangles at $\omega=\tfrac14 \pi$.}}
\label{Fig:SU(3)_nonSUSY_migration} 
\vspace{-0.2cm}
\end{figure*}

The last two solutions correspond to non-supersymmetric and $\textrm{SU}(3)$-invariant critical points. In the standard choice of $\omega=0$, the two solutions lie in the boundary of the moduli space and hence become unphysical. When $\omega$ starts running, the two critical points appear and flow again towards the boundary at $\omega=\pi/4$. Therefore, these critical points only exist in the new version of the $\textrm{SO}(8)$-gauged maximal supergravity. As a remark, the value of the CC as a function of $\omega$ does not peak at the special value $\omega=\pi/8$ as in the previous cases, but it is slightly shifted. At this value, the locations of the two critical points are
\be
\begin{array}{lclc}
z=-0.225 + i\, 0.306  & \hspace{5mm} , \hspace{5mm} &  \zeta_{12}=0.368-i\, 0.295 & , \\[2mm]
z=-0.306 + i\, 0.225  & \hspace{5mm} , \hspace{5mm} &  \zeta_{12}=-0.295+i\, 0.368 & ,
\end{array}
\ee
thus being related by (\ref{transf_sym}) and producing the same potential energy $V_{0}=-10.237$. The migration of the critical points  in field space as well as the potential  energy as a function of $\omega$ is depicted in figure~\ref{Fig:SU(3)_nonSUSY_migration}.

\begin{figure}[ht!]
\begin{tabular}{cc}
\includegraphics[width=70mm]{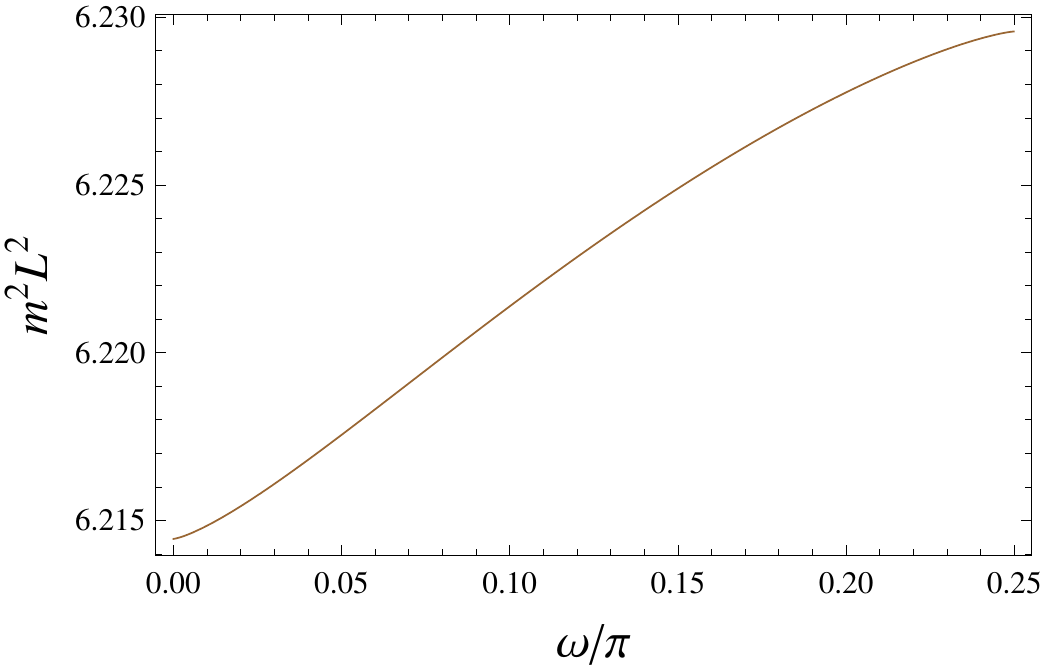}
&  
\includegraphics[width=70mm]{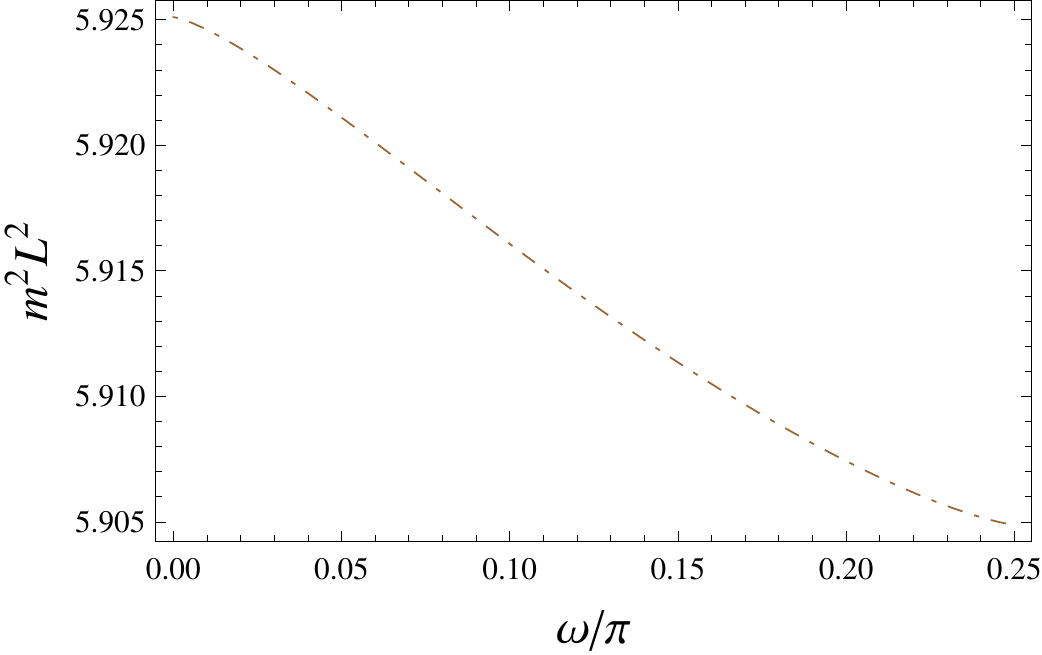}
\\
\includegraphics[width=70mm]{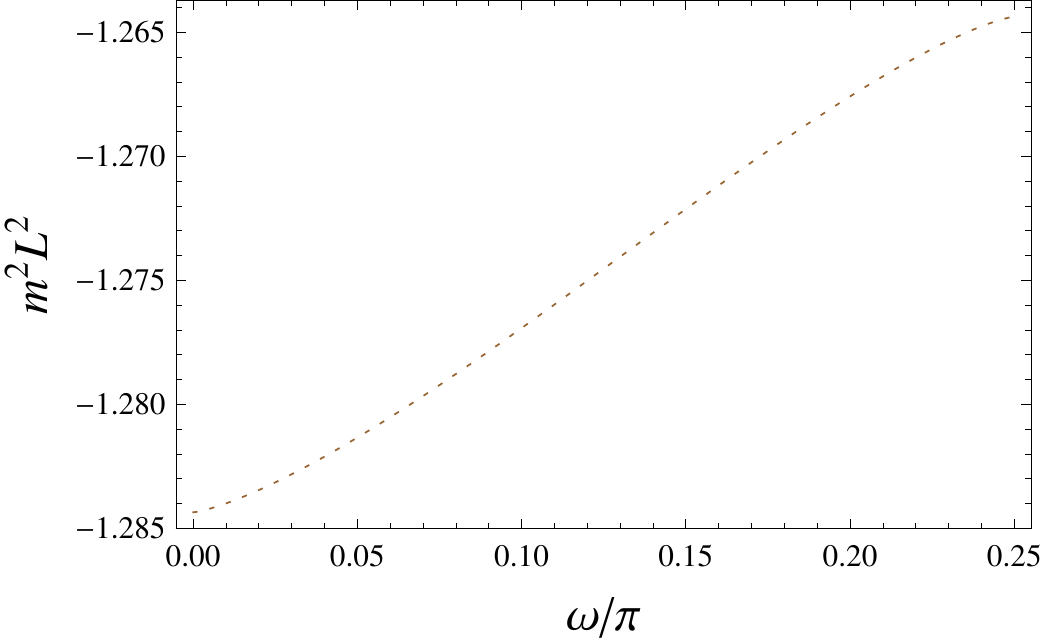}
&  
\includegraphics[width=70mm]{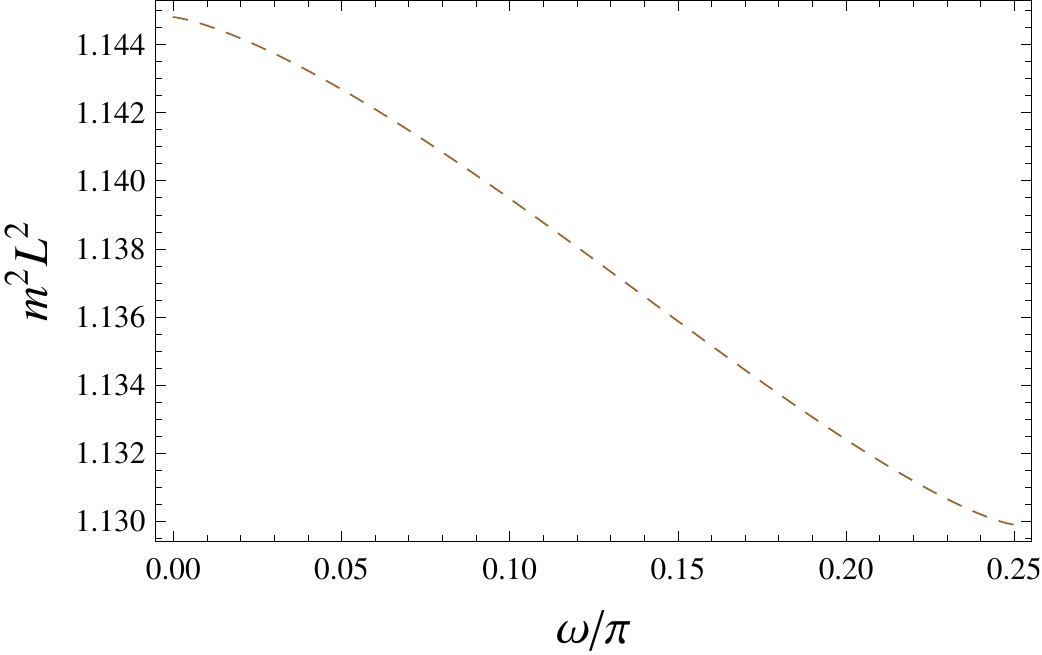} 
\end{tabular}
\begin{center}
 \includegraphics[width=70mm]{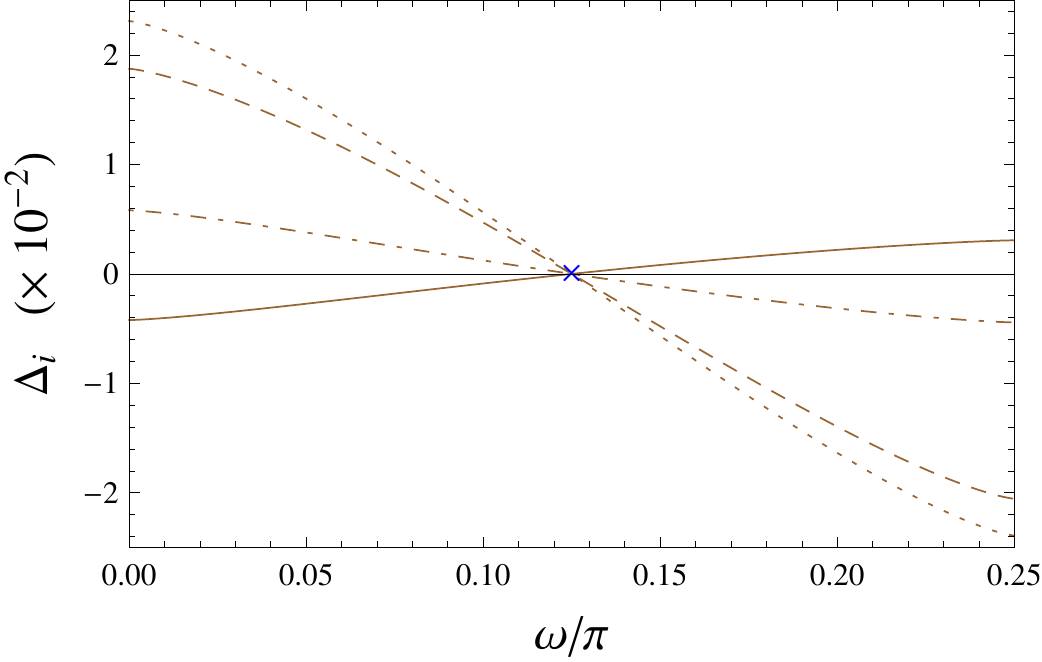}
 \end{center}
\caption{{\it Running of the eigenvalues of the mass matrix $m^2 L^2$ (first four figures) as a function of $\omega$ for one of the two non-supersymmtric and $\textrm{SU}(3)$-preserving solutions. We also give the relative change for the four masses (lower figure). The spectra for the other $\textrm{SU}(3)$-preserving critical point are mirror-symmetric around $\omega=\frac{\pi}{8}$.}}
\label{Fig:SU3eigenvalues}
\vspace{-0.2cm}
\end{figure}

Let us move to describe one of the most interesting and novel features of these non-supersymmetric critical points: they are the first examples of $\omega$-dependent masses in the SO$(8)$-gauged new maximal supergravity\footnote{The same behaviour has very recently been found in ref.\cite{Dall'Agata:2012sx} for (unstable) de Sitter solutions in the SO$(4,4)$ incarnation of new gauged supergravity and exploited to satisfy slow-roll conditions.}. To show this behaviour we have plotted in figure~\ref{Fig:SU3eigenvalues} the eigenvalues $m^2 L^2$ of the scalar mass matrix  as $\omega$ varies. It is worth mentioning here that the tachyonic field remains above the B.F. bound -- and hence stable within the $\textrm{SU}(3)$-invariant sector -- for any value of the $\omega$ parameter. However, full stability further requires the computation of all the $70$ scalar masses in maximal supergravity. We will come back to this issue in the next section. It is also interesting to note that the sum of four masses is $\omega$-independent and equals $12$.

The running of the mass eigenvalues with $\omega$ is a small effect which we will quantify in terms of their deviation from the value at $\omega=\pi/8$. Let us introduce the quantities $\Delta_{i}$ as 
\be
\Delta_{i}(\omega) \equiv \frac{m_{i}^2(\omega)-m_{i}^2(\pi/8) }{m_{i}^2(\pi/8)} \, L^2
\hspace{5mm} , \hspace{5mm} \textrm{ with } i=1,...,4 \ ,
\ee
which parameterise this deviation. The behaviour of these quantities is included in figure~\ref{Fig:SU3eigenvalues}, showing a relative running of the mass eigenvalues of around one per cent. It would be interesting to explore the field theory implications of this effect in case that a ``new" sphere reduction of 11d supergravity could be found.

\section{Spectra in the full $\cN=8$ theory}
\label{sec:N=8_spectra}

In order to derive the full mass spectra of the critical points found in the previous section, we now return to the full $\cN = 8$ theory and in particular all its 70 scalars and 28 electric and 28 magnetic vectors. To this end we will employ a method that was proposed in ref.~\cite{Dibitetto:2011gm} in half-maximal supergravity. It has been applied in maximal supergravity to classify the vacua supported by the scalars in the ${\bf 35}_{\rm v}$ \cite{DallAgata:2011aa} as well as the $\textrm{G}_2$-invariant vacua \cite{Borghese:2012qm}. In the present section we will extend the analysis of ref.~\cite{Borghese:2012qm} to the $\textrm{SU}(3)$ case.

\subsection*{Embedding tensor classification}

The crucial observation underlying this approach is that the embedding tensor, when dressed up with the scalar dependence to give the so-called T-tensor and evaluated at a critical point, is necessarily invariant under the  symmetries of that critical point. Moreover, as maximal supergravity has a homogenous scalar manifold, there is no loss of generality when assuming this point to be the origin. In other words, the classification of all critical points with a given symmetry is equivalent to the classification of all embedding tensors with that symmetry in the origin.

Our approach of restricting ourselves to the origin breaks the full $\textrm{E}_{7(7)}$ to its maximal compact part $\textrm{SU}(8)$. Hence everything should be translated into irreps of $\textrm{SU}(8)$, where the index $\,\mathcal{I}=1,...,8\,$ denotes the fundamental representation. The embedding tensor, transforming in the $\bf 912$ irrep of $\textrm{E}_{7(7)}$, then gives rise to the following pieces transforming in the $({\bf 36}\,\oplus\,{\bf 420})\,\oplus\,\textrm{c.c.}$ of $\textrm{SU}(8)$ 
\be
\begin{array}{lcclcclccl}
 \cA_1 & \equiv &  \cA^{\mathcal{I}\mathcal{J}} & \hspace{5mm}, &  &  &  \cA_2 & \equiv & \cA_{\mathcal{I}}{}^{\mathcal{J}\mathcal{K}\mathcal{L}} & ,
\end{array}
\ee
where $\cA^{\mathcal{I}\mathcal{J}} = \cA^{(\mathcal{I}\mathcal{J})}$, $\cA_{\mathcal{I}}{}^{\mathcal{J}\mathcal{K}\mathcal{L}} = \cA_{\mathcal{I}}{}^{[\mathcal{J}\mathcal{K}\mathcal{L}]}$ and $\cA_{\mathcal{I}}{}^{\mathcal{I}\mathcal{K}\mathcal{L}} = 0$.
In order to perform the truncation introduced in section~\ref{sec:N=2formul}, we first need to know how $\textrm{SU}(3)$ is embedded inside $\textrm{SU}(8)$. Such an embedding turns out to be defined by two possible chains of truncations
\be
\begin{array}{cccccccccccccc}
\textrm{E}_{7(7)} & \supset & \textrm{SU}(8) & \supset & \textrm{SO}(8) & \supset & \textrm{SO}(7) & \supset & \begin{array}{c}\nearrow\\
\searrow\end{array} & 
\hspace{-4mm}
\begin{array}{ccc}
& \textrm{G}_{2} &\\
 &  & \\
\textrm{SU}(4) & \supset & \textrm{U}(3)
\end{array} \hspace{-4mm} & 
\begin{array}{c}
\searrow\\
\nearrow
\end{array} & \supset & \textrm{SU}(3) & 
\end{array}
\nn
\ee
Of these, we will describe the $\textrm{G}_2$, the $\textrm{SU}(4)$ and the $\textrm{SU}(3)$ decompositions in more detail in what follows.

In addition to these bosonic symmetries, one can impose any number of supersymmetries. For instance, in the case of a single supersymmetry, one can identify a single entry of the $\textrm{SU}(8)$ index with the preserved supersymmetry; we will denote this by $1$. The requirement of $\cN = 1$ supersymmetry then reads
\be
\mathcal{N}=1 \,\,\, \colon
\hspace{8mm}
\mathcal{A}^{1 \times} = 0
\hspace{5mm} , \hspace{5mm}
{\mathcal{A}_{1}}^{\mathcal{I} \mathcal{J} \mathcal{K}} = 0  \ ,
\ee 
where $\times$ is anything but $1$.
Similarly, the requirement of two supersymmetries singles out two indices, $1, \hat 1$, that will correspond to the preserved Killing spinors. The algebraic conditions on the embedding tensor then read
\be
\mathcal{N}=2 \,\,\, \colon
\hspace{8mm}
\mathcal{A}^{1  \times} = 0
\hspace{5mm} , \hspace{5mm}
\mathcal{A}^{\hat 1  \times} = 0
\hspace{5mm} , \hspace{5mm}
{\mathcal{A}_{1}}^{\mathcal{I} \mathcal{J} \mathcal{K}} = 0
\hspace{5mm} , \hspace{5mm}
{\mathcal{A}_{\hat{1}}}^{\mathcal{I} \mathcal{J} \mathcal{K}} = 0  \ ,
\ee 
where $\times$ is anything but $1, \hat 1$.
Besides the above simplifications coming from requiring some preserved supersymmetry, one still has some local symmetry that can be used to bring the embedding tensor to a simpler form:
\begin{itemize}
\item[$i)$] In the case of $\mathcal{N}=1$ solutions, there is a $\textrm{U}(1) \times \textrm{U}(1) \times \textrm{U}(1) \subset \textrm{SU}(8)$ symmetry compatible with the supersymmetry conditions. 

\item[$ii)$] In the case of $\mathcal{N}=2$ solutions, there is a remaining $\textrm{U}(1) \times \textrm{U}(1) \times \textrm{SU}(2) \subset \textrm{SU}(8)$ symmetry. This is the subgroup of $\textrm{SU}(8)$ that commutes with $\textrm{SU}(3)$ and indeed arises as the maximal compact subgroup of the model in section 3. 

\end{itemize}
We will exploit these symmetries in what follows.

Given the $\textrm{G}_2$-,  $\textrm{SU}(4)$- and $\textrm{SU}(3)$-invariant ansatz for the embedding tensor components, we turn to an algebraic system of quadratic equations. These are first of all given by the quadratic constraints on the embedding tensor, that arise as consistency conditions for the gauging. They correspond to the $\textbf{133} \,\oplus\, \textbf{8645}$ irreps of $\textrm{E}_{7(7)}$. Secondly, we need to take the requirement that the origin is a critical point into account. This corresponds to the equations of motion for the scalar fields, which are represented by a ${\bf 70}_+$ irrep. The explicit form of both the quadratic constraints and the equations of motion in $\textrm{SU}(8)$ notation can be found in ref.~\cite{LeDiffon:2011wt}.

Based on techniques from algebraic geometry, in particular prime ideal decomposition and its implementation in the software \textsc{\,Singular\,} \cite{DGPS}, a huge set of critical points is revealed. However, in order to compare with the results in section 3, one has to keep in mind that any information about the gauging underlying a solution is lost in this approach. In particular, one fixes the residual symmetry (and possibly the amount of supersymmetry) to be preserved at the origin and then the set of possible gaugings compatible with this comes out after solving the quadratic constraints and equations of motion simultaneously. Thus, one would expect more solutions than those found in section 3 with other underlying gaugings not being the $\textrm{SO}(8)$ gauging. We will show that this is indeed what happens.

In what follows we will describe the Anti-de Sitter branches of solutions and their spectra in the three truncations. All of these correspond to solutions found in the previous section and hence are relevant for the $\textrm{SO}(8)$ gauging. Similar results on Minkowski branches for other gaugings can be found in appendix D.

\subsection*{The intermediate $\textrm{G}_{2}$ truncation}

Along the way through the $\textrm{G}_{2}$ trucation, the $\bf 8$ of $\textrm{SU}(8)$ goes into the ${\bf 8}_{\rm s}$ of $\textrm{SO}(8)$, then into the $\bf 8$ of $\textrm{SO}(7)$, and finally it splits into the ${\bf 1} \, \oplus \, {\bf 7}$ of $\textrm{G}_{2}$:
\be
\begin{array}{cccc}
\textrm{SU}(8) & \longrightarrow & \textrm{G}_{2} & , \\[2mm]
 \mathcal{I} & \longrightarrow & \left(1, \, m\right) & ,
\end{array}
\ee
where $m \, = \, 2,\,\dots,\,8$ labels the fundamental representation of $\textrm{G}_{2}$. The $\textrm{G}_{2}$-invariant components of the $\cA_1$ tensor read \cite{Borghese:2012qm}
\be
\label{G2A1}
 \begin{array}{lclc}
\cA^{11} = \alpha_1 & , &  \cA^{mn} = \alpha_2 \, \delta^{mn} & , 
 \end{array}
\ee
where $\alpha_1$ and $\alpha_2$ are arbitrary complex constants. The $\textrm{G}_{2}$-invariant components of the $\cA_2$ tensor are 
\be
\label{G2A2}
 \begin{array}{lclclclc} 
\cA_{1}{}^{mnp} = \beta_{1} \, \varphi^{mnp} & , & \cA_{m}{}^{1 np} = \beta_{2} \, \varphi_{m}{}^{np} & , & \cA_{m}{}^{npq} = \beta_{3} \, (*\varphi)_{m}{}^{npq} & , 
 \end{array}
\ee
where $\varphi$ and $*\varphi$ are, respectively, the $\textrm{G}_{2}$-invariant three-form and its dual four-form introduced in ref.~\cite{Borghese:2012qm} and $\textrm{G}_{2}$ indices are raised and lowered by means of $\delta^{mn}$ and its inverse. 

It is worth mentioning here that the $2+3$ complex constants introduced above are in agreement with the decomposition of the ${\bf 912}$ under \eqref{G2_trunc} yielding the ${\bf 2} \,\oplus\, {\bf 8}$ of $\textrm{SL}(2)$. As already pointed out in ref.~\cite{Borghese:2012qm}, an embedding tensor configuration of the form given in \eqref{G2A1} and \eqref{G2A2} further satisfies the requirement of $\textrm{SO}(7)_{\pm}$-invariance whenever $\alpha_{1}=\alpha_{2}$, $\beta_{2}=-\beta_{1}$ and $\beta_{3}=\pm \beta_{1}$. This agrees with the fact that the decomposition of the ${\bf 36} \,\oplus\, {\bf 420}$ contains two $\textrm{SO}(7)$-singlets and the same for the conjugate irreps. In order to go back to $\textrm{SO}(8)$-invariance instead, we further need all the $\beta$'s to be vanishing.

We find the following $\textrm{G}_2$-invariant branches of solutions \cite{Borghese:2012qm}.

\subsubsection*{$\bullet$ $\textrm{SO}(8)$-invariant vacua with $\cN = 8$}

This is the family of critical points with maximal $\textrm{SO}(8)$ residual symmetry and preserving maximal $\cN=8$ supersymmetry. The embedding tensor is given in this case by the simple expressions
\be
\mathcal{A}^{\mathcal{I}  \mathcal{J} } = \Lambda \, e^{i \theta}  \, \mathds{1}_{8}
\hspace{8mm} , \hspace{8mm} 
{\mathcal{A}_{\mathcal{I}}}^{\mathcal{J}  \mathcal{K} \mathcal{L} } = 0 \,\,\ ,
\ee
and produces an AdS$_4$ vacuum with energy $V_{0}=-6 \Lambda^2$. At this critical point, the scalar masses are given by
\be
\begin{array}{lclclclclc}
m^2 L^2 & = &  -2 \quad (\times 70)  & ,
\end{array}
\ee
and the vectors are all massless
\be
\begin{array}{lclclclclc}
m^2 L^2 & = &  0 \quad (\times 56) & .
\end{array}
\ee
In addition to 28 massless magnetic vectors, which will be present in all following branches as well, the 28 electric vectors are massless as well and generate the $\textrm{SO}(8)$ gauge group.

\subsubsection*{$\bullet$ $\textrm{G}_2$-invariant vacua with $\cN = 1$}

The non-vanishing embedding tensor components are
\be
\alpha_{1}=-2 \, \Lambda \, e^{-i 5 \theta}
\hspace{3mm} , \hspace{3mm}
\alpha_{2}=\sqrt{6} \, \Lambda \, e^{-i 5 \theta}
\hspace{3mm} , \hspace{3mm}
\beta_{2}=\sqrt{\frac{2}{3}} \, \Lambda \, e^{-i\theta}
\hspace{3mm} , \hspace{3mm}
\beta_{3}=\Lambda \, e^{i 3 \theta} \ ,
\ee
what fixes the vacuum energy to $V_{0}=-24 \Lambda^2$. The scalar spectrum consists of the following masses 
\be
\begin{array}{cccccccc}
\label{ScalarMassG2N1}
m^2 L^2 & = & 4 \pm \sqrt{6} \quad (\times 1) & , & -\frac{1}{6} \left(11 \pm \sqrt{6} \right)  \quad (\times 27) & , & 0 \quad (\times 14) & , \\
\Delta & = & \frac{1}{2} \left( 3 \pm 1 + 2 \sqrt{6} \right) & , & 2 \mp \frac{1}{\sqrt{6}}  & , & \text{unphysical} & , 
\end{array}
\ee
whereas the vector masses read
\be
\begin{array}{lclclclclclc}
\label{VectorMassG2N1}
m^2 L^2 & = &  0 \quad (\times 42) & , & \dfrac{1}{2} (3 \pm \sqrt{6}) \quad (\times 7)  & .
\end{array}
\ee
One observes that there are $14$ physical massless vectors associated to the $\textrm{G}_{2}$ residual symmetry while $14$ vectors acquire mass by eating up $14$ unphysical massless scalars. The list of $\textrm{OSp}(1|4)$ supermultiplets consists of a supergravity, a $\bf 7$ of massive gravitini, a $\bf 14$ of massless vectors and a ${\bf 1} \oplus \bf 27$ of chiral multiplets.

\subsubsection*{$\bullet$ $\textrm{SO}(7)_{\pm}$-invariant vacua with $\cN = 0$}

The associated non-vanishing embedding tensor parameters are given by 
\be
\alpha_{1} = \alpha_{2} = 3 \, \Lambda \,  e^{-i 3 \theta}
\hspace{5mm} , \hspace{5mm}
\beta_{1} = -\beta_{2} = \pm \beta_{3} = - \, \Lambda \, e^{i \theta} \ ,
\ee
producing a value for the energy of $V_{0}=-40 \Lambda^{2}$. The scalar masses and the vector masses are given by
\be
\begin{array}{ccccccccccccc}
m^2 L^2 & = & 0 \quad (\times 7) & , & 6  \quad (\times 1) & , & -\dfrac{6}{5} \quad (\times 35) & , & -\dfrac{12}{5}  \quad (\times 27)  & ,
\end{array}
\ee
and
\be
\begin{array}{cccccccccccccccc}
m^2 L^2 & = &  0 \quad (\times 49) & , & \dfrac{12}{5} \quad (\times 7)  & ,
\end{array}
\ee
respectively. As expected due to the residual symmetry, there are $21$ physical massless vectors.

\subsubsection*{$\bullet$ $\textrm{G}_{2}$-invariant vacua with $\cN = 0$}

The non-vanishing components of the embedding tensor now read
\be
\alpha_{1} = \sqrt{3} \, \Lambda \, e^{-i 3 \theta}
\hspace{5mm} , \hspace{5mm}
\alpha_{2} = -  \Lambda \, e^{-i 3 \theta}
\hspace{5mm} , \hspace{5mm}
\beta_{1} = \Lambda \, e^{i \theta}
\hspace{5mm} , \hspace{5mm}
\beta_{2} = \frac{1}{\sqrt{3}} \, \Lambda \,  e^{i \theta} \ ,
\ee
and give rise to $V_{0}=-4 \Lambda^2$. The masses for the scalars are given by 
\be
\begin{array}{ccccccccccccc}
m^2 L^2 & = & 0 \quad (\times 14) & , & 6  \quad (\times 2) & , & -1 \quad (\times 54)  & , 
\end{array}
\ee
and for the vectors by
\be
\begin{array}{cccccccccccccccc}
m^2 L^2 & = &  0 \quad (\times 42) & , & 3 \quad (\times 14)  & .
\end{array}
\ee
This time there are $14$ physical massless vectors associated to the residual symmetry group.

\subsection*{The intermediate $\textrm{SU}(4)$ truncation}

Let us now consider the route through the $\textrm{SU}(4)$ truncation. In the above diagram we have shown that the $\textrm{SU}(4)$-invariant sector of maximal supergravity also contains the $\textrm{SO}(7)$-invariant one in analogy with the $\textrm{G}_{2}$-truncated sector. Nevertheless, in order to parametrise the embedding tensor in a simpler way, we will rather choose to follow the chain\footnote{In terms of counting of degrees of freedom, the results are completely independent of the route chosen to embed $\textrm{SU}(4)$ inside $\textrm{SU}(8)$.}
\be
\label{chain:SU4_trunc}
\begin{array}{lclclc}
\textrm{SU}(8) & \supset & \textrm{U}(1)_{S} \,\times \, \textrm{SU}(4)_{1} \,\times \,\textrm{SU}(4)_{2} & \supset & \textrm{U}(1)_{S} \,\times \, \textrm{SU}(4)_{\textrm{diag}} & .
\end{array}
\ee
The $\textrm{SU}(4)_{\textrm{diag}}$ with respect to which we are truncating is obtained in the last step by anti-identifying the two $\textrm{SU}(4)$ factors above through $\textbf{4}_{1}\,\equiv\,\overline{\textbf{4}}_{2}$. In the present case the $\bf 8$ of $\textrm{SU}(8)$ branches into ${\bf 4}  \,\oplus\, {\bf \bar 4}$ of $\textrm{SU}(4)$. Therefore we use the following decomposition of the fundamental $\textrm{SU}(8)$ indices:
  \be
  \begin{array}{lclc}
   \mathcal{I} & \longrightarrow & ( \, i=1,...,4  \,\, , \,\, \hat{i}=5,...,8 \,) & ,
  \end{array}
 \ee
where $i$ and ${\hat i}$ denote the fundamental indices of the two different $\textrm{SU}(4)$'s. The anti-identification introduced in \eqref{chain:SU4_trunc} leads to the following new $\textrm{SU}(4)$-invariant tensors
\be
\begin{array}{cccccc}
\delta^{i\hat{j}} & , & {\eps_{i}}^{\hat{j}\hat{k}\hat{l}} & , & {\eps_{\hat{i}}}^{jkl} & .
\end{array}
\ee
By making use of these, one can construct the following set of invariant embedding tensor components
\be
\label{SU4A1}
 \begin{array}{lclclc}
\cA^{i \hat{j}} & = &  \cA^{\hat{j} i} & = & \gamma \, \delta^{i\hat{j}} & ,
 \end{array}
\ee
for what regards $\cA_1$, together with the set of $\cA_2$ components
\be
\label{SU4A2}
 \begin{array}{lclclclc} 
\cA_{i}{}^{jk \hat{l}} = \delta_{1} \, \delta^{[j}_{i} \, \delta^{k] \hat{l}}  & , & \cA_{\hat{i}}{}^{jkl} = \delta_{2} \, {\eps_{\hat{i}}}^{jkl} & , & \cA_{\hat i}{}^{\hat{j} \hat{k} l } = \delta_{3} \, \delta^{[\hat{j}}_{\hat{i}} \, \delta^{\hat{k}] l}  & ,  &  \cA_{i}{}^{\hat{j}\hat{k}\hat{l}} = \delta_{4} \, {\eps_{i}}^{\hat{j}\hat{k}\hat{l}} & ,
 \end{array}
\ee
where the condition $\delta_{1} = \delta_{3}$ is required by the tracelessness of the ${\bf 420}$. Due to this, the number of independent complex parameters for the $\textrm{SU}(4)$-invariant embedding tensor reduces to $4$. This agrees with the decomposition of the ${\bf 912}$ under 
\be
\begin{array}{lclclcl}
\textrm{E}_{7(7)} & \supset & \textrm{SL}(8) & \supset & \mathbb{R}^{+}_{T} \, \times \, \textrm{SL}(2)_{S} \, \times \, \textrm{SL}(6) & \supset & \mathbb{R}^{+}_{T} \, \times \, \textrm{SL}(2)_{S} \, \times \, \textrm{SU}(4) \,,
\end{array}
\ee
giving rise to $\textrm{SU}(4)$-singlets in the ${\bf 1} \,\oplus\, {\bf 1} \,\oplus\, {\bf 3}  \,\oplus\, {\bf 3}$ of $\textrm{SL}(2)_{S}$.

In this truncation, we only find a single branch of AdS solutions apart from the maximally supersymmetric $\textrm{SO}(8)$-invariant and the non-supersymmetric $\textrm{SO}(7)$-invariant ones.

\subsubsection*{$\bullet$ $\textrm{SU}(4)$-invariant vacua with $\cN = 0$}

 The embedding tensor parameters are given by
\be
\gamma = \mp \frac{3 \Lambda}{2 \sqrt{2}} \, e^{ i3\theta}
\hspace{5mm} , \hspace{5mm}
\delta_{1} = \delta_{3} = \mp \frac{\Lambda}{\sqrt{2}} \, e^{- i \theta} 
\hspace{5mm} , \hspace{5mm}
\delta_{2} = - \delta_{4} = -  \Lambda \, e^{- i \theta} \ ,
\ee
and produce a vacuum energy of $V_{0}=-4 \Lambda^2$. The scalar masses take the values
\be
\begin{array}{ccccccccccccc}
m^2 L^2 & = & 0 \quad (\times 28) & , & 6  \quad (\times 2) & , & -3 \quad (\times 20) & , & -\dfrac{3}{4} \quad (\times 20) & ,
\end{array}
\ee
whereas the vector masses are
\be
\begin{array}{cccccccccccccccc}
m^2 L^2 & = &  0 \quad (\times 43) & , & \dfrac{9}{4} \quad (\times 12) & , & 6 \quad (\times 1)  & .
\end{array}
\ee
Among these, there are $15$ physical massless vectors associated to the residual symmetry. The same scalar spectrum was found in ref.~\cite{DallAgata:2011aa} and associated to an $\textrm{SO}(8)$ gauging among other possibilities.

\subsection*{The $\textrm{SU}(3)$ truncation}

We now turn to the case of $\textrm{SU}(3)$-invariance. In this case one has the natural index splitting $i=(1,a)$ and $\hat{i}=(\hat{1},\hat{a})$, and can build the following $\textrm{SU}(3)$-invariant components for the $\cA_1$ tensor
\be
\label{SU3A1}
 \begin{array}{lclclclc}
\cA^{1 \hat 1} = \cA^{\hat{1} 1} = \lambda_1 & \,\,,\,\, &  \cA^{a \hat b} = \cA^{\hat{b} a} = \lambda_2 \, \delta^{a \hat b} & \,\,,\,\, & \cA^{11} = \lambda_3 & \,\,,\,\, &  \cA^{\hat 1 \hat 1} = \lambda_4 & .
 \end{array}
\ee
The first two components happen to enjoy two additional Abelian symmetries, which we will refer to as $\textrm{U}(1)^{(1),(2)}$. The corresponding charges are listed in table~\ref{Table:U(1)U}. 

\begin{table}[ht]
\renewcommand{\arraystretch}{1.15} 
\begin{center}
\begin{tabular}{|c||c|c|c|c|}
\hline 
 & $1$ & $a$ & $\hat 1$ & $\hat a$ \\[1mm]
\hline \hline
$\textrm{U}(1)^{(1)}$  & $+3$ & $+1$ & $-3$ & $-1$ \\[1mm]
\hline \hline
$\textrm{U}(1)^{(2)}$  & $+3$ & $-1$ & $-3$ & $+1$ \\[1mm]
\hline
\end{tabular}
\caption{{\it The charges of the $\,( 1 \,,\, a \,,\, \hat{1} \,,\,\hat{a} )\,$ indices under the two relevant $\textrm{U}(1)$ embeddings inside $\textrm{SU}(8)$. These are related via an  $a \leftrightarrow \hat{a}$ interchange.}}
\label{Table:U(1)U} 
\end{center}
\vspace{-.5cm}
\end{table}

Similarly, the $\cA_2$ tensor is parametrised by the following $\textrm{SU}(3)$-invariant components
\be
\label{U3A2}
 \hspace{-1mm}
 \begin{array}{lclclclc}
  \cA_1{}^{bcd} = \mu_1 \, \epsilon^{bcd} & , &  \cA_{\hat 1}{}^{\hat b \hat c \hat d} = \mu_2 \, \epsilon^{\hat b \hat c \hat d} & , & \cA_{\hat a}{}^{\hat 1 b c} = \mu_3 \, \epsilon_{\hat a}{}^{bc} & , & \cA_{a}{}^{1 \hat b \hat c} = \mu_4 \, \epsilon_{a}{}^{\hat b \hat c} & , \\[2mm]  
  \cA_1{}^{1 a \hat b} = \mu_5 \, \delta^{a \hat b} & , & \cA_{\hat 1}{}^{\hat 1 a \hat b} = \mu_6 \, \delta^{a \hat b} & , & \cA_a{}^{b c \hat d} = \mu_7 \, \delta_a^{[b} \delta^{c] \hat d} & , & \cA_{\hat a}{}^{\hat b \hat c d} = \mu_{8} \, \delta_{\hat a}^{[\hat b} \delta^{\hat c] d} & , \\[2mm]
 & & \cA_{a}{}^{b 1 \hat 1} = \mu_9 \, \delta_{a}^{b} & , & \cA_{\hat a}{}^{\hat b 1 \hat 1} = \mu_{10} \, \delta_{\hat a}^{\hat b} & , & & 
 \end{array}
\ee 
which are  $\textrm{U}(1)^{(1)}$-invariant, plus the following additional ones being purely $\textrm{SU}(3)$-invariant
\be
\label{SU3A2}
 \hspace{-1mm}
 \begin{array}{lclclclc} 
\cA_{a}{}^{\hat 1 \hat b \hat c} = \mu_{11} \, \epsilon_{a}{}^{\hat b \hat c} & , & \cA_{1}{}^{\hat b \hat c \hat d} = \mu_{12} \, \epsilon^{\hat b \hat c \hat d} & , & \cA_{\hat a}{}^{1 bc} = \mu_{13} \,   \epsilon_{\hat a}{}^{bc}  & , & \cA_{\hat 1}{}^{bcd} = \mu_{14} \, \epsilon^{bcd} & , \\[2mm]
& & \cA_{1}{}^{\hat 1 a \hat b} = \mu_{15} \, \delta^{a \hat b} & , & \cA_{\hat 1}{}^{1 a \hat b} = \mu_{16} \, \delta^{a \hat b} & , & & 
 \end{array}
\ee
where $\epsilon_{a}{}^{\hat b \hat c} \equiv \delta_{a \hat d} \, \epsilon^{\hat b \hat c \hat d}$, etc. An analogous reasoning can be done in terms of $\textrm{U}(1)^{(2)}$-invariant components. On the other hand, the tracelessness of the $\bf 420$ irrep corresponds to the following linear constraints
\be
 \begin{array}{lccclc}
  \mu_{5} \,+\, \mu_{6} \,+\, \mu_{7} \,-\, \mu_{8} \,=\, 0 &  & \textrm{and} &  & \mu_{9} \,+\, \mu_{10} \,=\, 0 & .
 \end{array}
\ee
Note that the $20$ complex parameters introduced in \eqref{SU3A1}-\eqref{SU3A2} subject to the above linear contraints exactly give rise to the $36$ real deformation parameters that are present in the decomposition
\be
\label{SU3ET}
\begin{array}{lclcclcl}
{\bf 912} & \rightarrow & ({\bf 2},{\bf 1}) \, \oplus \, ({\bf 2},{\bf 1}) \, \oplus \, ({\bf 4},{\bf 8}) 
\end{array} 
\ee
for the embedding tensor as irreps of $\,\textrm{SL}(2) \times \textrm{SU}(2,1)$. Here we have suppressed $\textrm{SU}(3)$ non-singlet terms.

We find the following additional branches of solutions in this truncation.

\subsubsection*{$\bullet$ $(\textrm{SU}(3) \times \textrm{U}(1)^{(1)})$-invariant vacua with $\cN = 2$}

This family of AdS$_4$ solutions has energy  $V_{0}=-\frac{3\Lambda^2}{8}$ and non-vanishing embedding tensor parameters given by 
\be
\label{ETWarner1}
\begin{array}{lclclc}
\lambda_{1} = -\frac{\Lambda}{4} \, e^{i\theta} & , & \lambda_{2} = - \frac{\Lambda}{3} \, e^{-i 7 \theta}  & , & \mu_{7} = \mu_{8} =\frac{\Lambda}{3}  \, e^{i 3 \theta} & , \\[2mm]
\mu_{3} = \mp \frac{\Lambda}{2\sqrt{3}} \, e^{i 4 \theta} & , & \mu_{4} = \pm \frac{\Lambda}{2\sqrt{3}}  & , & \mu_{9} = -\mu_{10} = \frac{\Lambda}{12}  \, e^{i \theta} & ,
\end{array}
\ee
%
%
where $\Lambda$ again sets the scale of the CC. It is straightforward to check that the associated set of embedding tensor components\footnote{In addition there is a completely analogous branch of solutions with $\textrm{SU}(3) \times \textrm{U}(1)^{(2)}$ invariance producing the same CC and mass spectra.} is left invariant by $\textrm{U}(1)^{(1)}$ and not by $\textrm{U}(1)^{(2)}$. 

The scalar mass spectrum at this family of critical points is given by
\be
\begin{array}{cccccccc}
\label{ScalarMassWarner1}
m^2 L^2 & = & 3 \pm \sqrt{17} \quad (\times 1) & , & 2 \quad (\times 3) & , & 0 \quad (\times 19) & , \\[2mm]
& & -\frac{14}{9} \quad (\times 18) & , & -2 \quad (\times 16) & , & -\frac{20}{9} \quad (\times 12)  & , \\[4mm]
\Delta & = & \frac{1}{2} \left( 3 \pm 2 + \sqrt{17} \right) & , &  \frac{1}{2} \left( 3 + \sqrt{17} \right) & , & \text{unphysical} & ,  \\[2mm]
& & \frac{7}{3} & , & 2 & , & \frac{5}{3} & , 
\end{array}
\ee
whereas the vector masses read
\be
\begin{array}{lclclclclclc}
\label{VectorMassWarner1}
m^2 L^2 & = &  0 \quad (\times 37) & , & \frac{4}{9} \quad (\times 12) & , & \frac{28}{9} \quad (\times 6)  & , & 4 \quad (\times 1) & .
\end{array}
\ee
This time there are $9$ physical massless vectors reflecting the residual symmetry group.  This family of solutions contains the one already found in ref.~\cite{Bobev:2010ib} by setting the AdS scale to  $\Lambda^2=12 \, \sqrt{3}$. For values of $\omega \neq 0$ it corresponds to the $\cN = 2$ solutions of section 3. The $\textrm{OSp}(2|4)$ supermultiplets are, in addition to those of section 3 as well as the supergravity multiplet, an $\bf 8$ of massless vector multiplets, ${\bf 3} \oplus \bar {\bf 3}$ short gravitino multiplets and a $\bf 6$ of hypermultiplets \cite{Klebanov:2008vq}.

\subsubsection*{$\bullet$ $\textrm{SU}(3)$-invariant vacua with $\cN = 1$}

The last family of AdS$_{4}$ solutions still preserving some supersymmetry is given by the following embedding tensor parameters
\be
\begin{array}{lclclclc}
\lambda_{2} = \frac{2}{3}\Lambda \, e^{ i4\theta} & , & \lambda_{3} = \pm \frac{\Lambda}{2} \, e^{i\theta} & , & \lambda_{4} = \Lambda & , & \mu_{2} = \frac{\sqrt{3}}{2}\Lambda \, e^{ i\theta} & , \\[2mm]
\mu_{3} = \frac{\Lambda}{\sqrt{3}} \, e^{- i3\theta} & , & \mu_{7} = \mu_{8} = \mp \frac{\Lambda}{3} \, e^{- i\theta} & , & \mu_{9} = -\mu_{10} =  \frac{\Lambda}{3} \, e^{- i2\theta} & , & \mu_{11} = -\frac{\Lambda}{\sqrt{3}} \, e^{- i3\theta} & , \\[2mm]
\mu_{13} = \mp \frac{\Lambda}{2\sqrt{3}} & , & \mu_{14} = - \frac{\sqrt{3}}{2}\Lambda \, e^{ i\theta} & , & \mu_{16} = \frac{\Lambda}{2} \, e^{ i2\theta} & , &  & 
\end{array}
\ee
%
%
%
%
and has a vacuum energy $V_{0}=-\frac{3 \Lambda^2}{2}$. Up to our knowledge, this is a new family of solutions of maximal gauged supergravity and has a representative element in the $\textrm{SO}(8)$ gauging (see section 3). 

The mass spectrum at any critical point in this family is given by
\be
\begin{array}{ccccccccccccc}
\label{ScalarMassSU3N1}
m^2 L^2 & = & 4 \pm \sqrt{6} \quad (\times 2) & , & -\frac{20}{9}  \quad (\times 12) & , & -2 \quad (\times 8) & ,\\[2mm] 
&  & -\frac{8}{9}  \quad (\times 12) & , & \frac{7}{9}  \quad (\times 6) & , & 0 \quad (\times 28) & , \\[4mm]
\Delta  & = & \frac{1}{2} \left( 3 \pm 1 + 2 \sqrt{6} \right) & , & \frac{5}{3} & , & 2 & ,\\[2mm] 
&  &\frac{8}{3} & , & \frac{1}{6} \left( 9 + \sqrt{109} \right) & , & 3 \, (\times 8) + \text{unphysical} & ,
\end{array}
\ee
whereas the vector masses read
\be
\begin{array}{cccccccccccccccc}
\label{VectorMassSU3N1}
m^2 L^2 & = &  0 \quad (\times 36) & , & \frac{4}{9} \quad (\times 6)  & , & 2 \quad (\times 1) & , \\[4mm]
&   & \frac{25}{9} \quad (\times 6)  & , & \frac{28}{9} \quad (\times 6)  & , & 6 \quad (\times 1)  & .
\end{array}
\ee
In this case one finds $8$ physical massless vectors associated to the $\textrm{SU}(3)$ residual symmetry. Again this corresponds to the family of points encountered in section 3. The additional supermultiplets in this case are ${\bf 3} \oplus {\bf \bar 3}$ massive gravitini, $\bf 8$ massless vectors, ${\bf 3} \oplus {\bf \bar 3}$ massive vectors, ${\bf 6} \oplus {\bf \bar 6}\oplus{\bf 8}$ chiral supermultiplets.

\subsubsection*{$\bullet$ $\textrm{SU}(3)$-invariant vacua with $\cN = 0$}

The family of non-supersymmetric solutions preserving $\textrm{SU}(3)$ is particularly hard to analyse using the embedding tensor classification approach. The reason is that the lack of residual supersymmetry reduces the set of embedding tensor simplifications to the local symmetry group. The resulting algebraic system consisting of quadratic constraints and equations of motion becomes very complex and we fail in decomposing it using algebraic geometry techniques.

Following the observation that all the previous $\theta$-dependent families of embedding tensor configurations happen to contain a real representative for $\theta=0$, we have explored this simplified setup for a (partial) classification of non-supersymmetric solutions at the $\theta=0$ point. This can be exhaustively analysed by using algebraic geometry techniques and, remarkably, we find the $\theta=0$ representative of all the solutions discussed so far in the paper (and in appendix D), and no more\footnote{Of course, this does not exclude the existence of other $\theta$-dependent families of non-supersymmetric solutions. There might be families not containing a real representative at $\theta=0$ and hence are missed in the real embedding tensor simplification.}. 

Let us focus on the non-supersymmetric and $\textrm{SU}(3)$-preserving (real) embedding tensor configuration obtained at $\theta=0$. The scalar mass spectrum at this point is given by
\be
\begin{array}{ccccccccccccc}
m^2 L^2 & = & 6.241  \quad (\times 1) & , & 5.888  \quad (\times 1) & , & -1.237 \quad (\times 1) & , &  1.107 \quad (\times 1) & ,\\[4mm] 
    &  & -1.411  \quad (\times 12) & , & -1.100  \quad (\times 18) & , & -1.082 \quad (\times 8) & , & -0.554 \quad (\times 8) & , \\[4mm]
&  & 0 \quad (\times 20) & ,
\end{array}
\ee
where the masses in the first line correspond to the $\textrm{SU}(3)$-invariant scalars. By comparing with figure~\ref{Fig:SU3eigenvalues}, one observes that such four masses do not lie inside the range allowed by the $\textrm{SO}(8)$ gauging, although there is a tiny difference of order $1\%$.  This reflects the fact that, even though it does not correspond to an $\textrm{SO}(8)$ gauging (indeed, we have checked that it corresponds to and SO$(7,1)$ gauging), the family of non-supersymmetric and $\textrm{SU}(3)$-invariant vacua is indeed captured at the $\theta=0$ point\footnote{We have also checked the presence of $8$ physical massless vectors associated to an $\textrm{SU}(3)$ residual symmetry group.}. A further confirmation is the fact that the four singlet masses again add up to $12$, in concordance with the previous section. Finally, assuming differences of the same order for the rest of the scalar masses, these non-supersymmetric solutions are very likely to be totally stable with respect to all the scalars in maximal supergravity. We would like to come back to this issue in the future.


\section{The $\textrm{G}_2$-invariant sector} \label{sec:FurtherTrunc}

Finally we turn to the $\textrm{G}_2$-invariant truncation with $\cN =1$ supersymmetry. In this sector we identify the two complex fields $z = \zeta_{12}$ while setting the vectors equal to zero. This theory can be obtained as a truncation of maximal supergravity with respect to a compact $\textrm{G}_2$ subgroup of $\textrm{E}_{7(7)}$ by following 
\be
\label{G2_trunc}
\begin{array}{cccl}
\textrm{E}_{7(7)} & \supset &  \textrm{SL}(2) \, \times \, \textrm{G}_{2} & .
\end{array}
\ee
The branching of the fundamental representation reveals the absence of $\textrm{G}_2$ singlets and hence the absence of vectors in this theory. The only invariant scalars turn out to span the $\textrm{SL}(2)/\textrm{SO}(2)$ coset parametrised by the complex scalar $z = \zeta_{12}$. Due to the fact that this truncation preserves $\cN = 1$ supersymmetry, an alternative way to describe this subsector of maximal supergravity is in terms of a real K\"ahler potential and holomorphic superpotential. These turn out to be given by
\begin{align} \label{N=1 in Warner variables}
 & \cK = -7 \, \ln \left[ -1 + \frac{1}{1+z} + \frac{1}{1+ \bar{z}} \right] \, , \;
 \cW = \frac{\sqrt{2} \left[ (1 + 7 \, z^{4}) \, e^{i \, \omega} + (7 \, z^{3} + z^{7}) \, e^{- i \, \omega} \right]}{(1 + z)^{7}} \ , 
\end{align}
out of which one can construct the scalar potential via
 \begin{align}
V = e^{\cK} \left[ \cK^{z \bar{z}} \, (\cD_{z} \cW) \, (\cD_{\bar{z}} \, \overline{\cW}) - 3 \, \cW \, \overline{\cW} \right] \ . \label{N=1potential}
 \end{align}
Performing the following holomorphic change of variable
\begin{align} \label{DiskToUHF}
z = \frac{i-S}{i+S} \ ,
\end{align}
from a parametrisation of the unit disc ($z$) to that of the half plane ($S$), the K\"ahler potential and the holomorphic superpotential read
 \be
 \begin{array}{ccll}
\cK & = & - 7 \, \ln \left[ - \dfrac{i}{2} \, ( S - \bar{S} ) \right] & , \\[2mm]
\cW & = & \dfrac{1}{4 \sqrt{2}} \left[ \left( 1 + 7 \, S^{4} \right) \cos \omega - \left( 7 \, S^{3} + S^{7} \right) \sin \omega \right] & .
\end{array}
 \ee
with $S = \chi + i \, e^{- 2 \phi / 7}$. The dilaton of this parametrisation can be identified with the $\textrm{SO}(7)_+$-invariant scalar. Indeed the scalar potential for this field coincides with that of ref.~\cite{Dall'Agata:2012bb}. The dependence of \eqref{N=1potential} on the other $\cN = 1$ field, the $\textrm{G}_2$-invariant pseudo-scalar $\chi$, then follows from holomorphicity of the superpotential. The two-field scalar potential is subtly different from that of ref.~\cite{Dall'Agata:2012bb} but agrees on the value of the CC in its critical points for all the values of $\omega$. Due to this highly non-trivial confirmation, we expect it to be related via a field redefinition\footnote{The absence of scalar kinetic terms in \cite{Dall'Agata:2012bb} complicates the explicit construction of this redefinition.}.


\section{Outlook} \label{sec:outlook}

In the present paper we have constructed and investigated the $\textrm{SU}(3)$-invariant sector of new maximal supergravity. This theory is a one-parameter extension of the old maximal $\textrm{SO}(8)$-gauged supergravity; the phase $\omega$ delineates the linear combination of electric and magnetic vectors that are employed in the gauging. We have demonstrated the modifications to the theory due to $\omega$, both in the superpotential as well as in the canonical formulation of the $\cN =2$ truncation to the $\textrm{SU}(3)$-invariant sector. In addition we have analysed the vacuum structure in detail. Our results indicate a number of novel features for new maximal supergravity as contrasted to the old theory. When moving from $\omega = 0$ to the bulk of the parameter space, the number of critical points doubles. This is illustrated in table \ref{Table:SU(3)2}. We have found that some of the long-known $\omega=0$ points partner for $\omega \neq 0$ with new points with the same spectrum. Further, we have also found altogether new critical points with no $\omega \neq 0$ analog. This is the case of our new $\cN=1$ and non-supersymmetryic SU(3) points. In the latter case, the observation that the number of points in this case increases with two when moving from $\omega = 0$ to $\omega \neq 0$ is intimately related to the ``twin peak'' structure of the cosmological constants of this branch, as plotted in figure \ref{Fig:SU(3)_nonSUSY_migration}.

In addition to the location and number of critical points, we have investigated their mass spectra for both the scalars and the vector fields. In concordance with previous results, these spectra turn out to be $\omega$-independent in most cases. The unique exception that we encountered within the $\textrm{SO}(8)$-gauged theory is the least symmetric branch, preserving $\cN = 0$ and only $\textrm{SU}(3)$. In this case, the mass spectra in fact turns out to be $\omega$-dependent, albeit very weakly. The variation of the scalar masses as a function of $\omega$ can be found in figure \ref{Fig:SU3eigenvalues}. 

Furthermore, we have stressed that another physical difference between different values of $\omega$ are the cosmological constants of the different branches. In units where the maximally supersymmetric vacuum has $V_0 = -6$, the CC of all other vacua are $\omega$-dependent. We have explicitly plotted this dependence for the different branches, and summarised the differences between $\omega = 0$ and $\pi/8$ in table \ref{Table:SU(3)2}. The ratios between the CC of different vacua have important holographic implications and, if connected via a holographic RG flow, this ratio should be reproduced from a calculation of the field theory's free energy $F$ at both ends of the flow. For $\omega=0$, this ratio has indeed been successfully reproduced from the field theory for the flow \cite{Ahn:2000mf,Corrado} between the $\cN=8$ and $\cN=2$ points \cite{F-theorem,Klebanov:2008vq,Benna:2008zy,Gabella:2011sg,Gabella:2012rc}. The same should happen for other values of $\omega$, if the new SO(8) gaugings of \cite{Dall'Agata:2012bb} are to have a field theory dual interpration. If ABJ \cite{ABJ} is the field theory dual to the $\omega = \pi/8$ gauging, as suggested in  \cite{Dall'Agata:2012bb}, then $F_\textrm{IR}/F_\textrm{UV}$ ought to reproduce the ratio $ V_{\cN = 8} /V_{\cN = 2}$ at $\omega = \pi/8$ ({\it i.e.}, minus six over (\ref{CCN=2})) between the CC constants at the $\cN=8$ and (one of the two) $\cN=2$ points. Similarly, if there is a field theory dual at $\omega=\pi/4$ then we can predict the same $F_\textrm{IR}/F_\textrm{UV}$ between the $\cN=8$ and $\cN=2$ points as in the original $\omega=0$ case. Table  \ref{Table:SU(3)2} clearly shows possible RG flow directions between the different critical points.

 The reason that we have restricted $\omega$ to the range from $0$ up to $\pi/4$ in the discussion throughout the paper is the following. The theory is invariant under a shift of $\omega$ with $\pi/2$ combined with an overall sign flip of the scalar fields. As the latter is a field redefinition, the parameter space of inequivalent theories is periodic in $\pi/2$. Moreover, replacing $\omega$ by $- \omega$ amounts to complex conjugating $z$ and $\zeta_{12}$. Under this operation, the superpotential $\cW = |\cW_\pm|$ and the vector kinetic term \eqref{gaugeKinPrime} are invariant while the topological term changes sign. As this is a parity operation, $\omega$ can be restricted to the range quoted above. Finally, while the number of critical points as well as their mass spectra and CC is also periodic in $\pi/4$, the embedding of their residual gauge symmetry is actually different at $\omega$ and $\omega+\pi/4$. For instance, while $\omega = 0$ gives rise to two $\textrm{SO}(7)_-$ and one $\textrm{SO}(7)_+$-invariant points, these numbers are interchanged at $\omega = \pi/4$, and similar for the embeddings of $\textrm{SU}(4)$ and $\textrm{SU}(3) \times \textrm{U}(1)$. It therefore remains to be seen whether these theories are physically equivalent.

At this point it is worthwhile to recap the evidence for the proposed superpotential. First of all, if the superpotential formulation extends from the old to the new theory, our proposal is the unique superpotential consistent with the various symmetries. Secondly, it is consistent with the canonical $\cN =2$ formulation. Thirdly, the cosmological constants of the $\textrm{G}_2$- and $\textrm{SO}(7)$-invariant critical points agree with \cite{Dall'Agata:2012bb}. Fourthly, the full scalar dependence on the two $\textrm{SO}(6)$-invariant dilatons coincides with that following from an $\cN = 8$ superpotential, as we demonstrate in appendix A. And last but not least, the mass spectra following from the $\cN =2$ scalar potential are consistent with those derived in section 4, which presents a complementary derivation of these. In particular, the results of section 4 do not depend in any way on the scalar potential of section 3. For these reasons we are confident that our scalar potential captures the correct dynamics. It would be interesting to construct it explicitly starting from the $\cN = 8$ scalar potential.

In addition to the construction and vacuum analysis of the $\textrm{SU}(3)$-invariant sector, our results confirm the interpretation of the $\omega$-phase as a symplectic rotation of electric and magnetic vectors and show how this rotation appears in a similar fashion in less supersymmetric theories. In particular, from the symplectically covariant formulation of section 2.3 it follows that the $\omega$-phase in $\cN = 2$ also corresponds to a rotation of electric and magnetic vectors, while leaving the scalar fields invariant. It is true that the R-symmetry of this theory includes an overall $\textrm{U}(1)$ transformation with the same action on the vector fields, but this transformation additionally acts on the scalar fields. The latter transformation is a symmetry and hence does not change the physics of the theory. Performing only the symplectic rotation on the vectors does induce a physical difference, as we have demonstrated.

We hope the present paper contributes to a further understanding of new maximal supergravity, in particular its vacuum structure, and look forward to interesting results on other open issues, including its higher-dimensional origin and holographic dual.



\section*{Acknowledgements}


We would like to thank Gianguido Dall'Agata, Jean-Pierre Derendinger, Bernard de Wit, Gianluca Inverso, Daniel Jafferis, Henning Samtleben, Mario Trigiante and Stefan Vandoren for very useful discussions. The research of AB and DR is supported by a VIDI grant from the Netherlands Organisation for Scientific Research (NWO). The work of GD is supported by the G\"oran Gustafsson Foundation. The work of AG is supported by the Swiss National Science Foundation. OV is supported in part by NWO under the VICI grant 680-47-603, and by the Spanish Government research grant FIS2008-01980.

\appendix

\section{The ${\bf 35}_{\rm v}$ scalars of maximal supergravity}
\label{N=8_SL8}

Maximal supergravity contains 70 scalar fields. For the $\textrm{SO}(8)$ gauging, these transform in two irreps of the gauge group, which can be chosen to be ${\bf 35}_{\rm v}$ and ${\bf 35}_{\rm s}$. The subscript label the inequivalent 35-dimensional irreps of $\textrm{SO}(8)$ which can be constructed as the symmetric tensor product of the vector and spinor irrep, respectively. The former are proper scalars while the latter are pseudo-scalars.

For one of these irreps, which we will take to be the ${\bf 35}_{\rm v}$, one can construct the full scalar potential from a superpotential. In particular, this irrep corresponds to the coset $\textrm{SL}(8)/ \textrm{SO}(8)$, and hence we have to restrict ourselves to gaugings withing $\textrm{SL}(8)$. The most general such gauge group (but not the most general gauging, as we will later see) is characterised by a symmetric matrix, which is often denoted by $Q_{ab}$. Without loss of generality it can be taken to be diagonal with entries equal to $0$ or $\pm 1$. The resulting gauge group in this case is $\textrm{CSO}(p,q,r)$, for $p$ positive, $q$ negative and $r$ vanishing entries. In particular, for the $\textrm{SO}(8)$ gauging one can take $Q_{ab}$ to be the identity.

Note that the vacuum structure for this truncation of the maximal theory has been exhaustively analysed in ref.~\cite{DallAgata:2011aa}. Subsequently, a group-theoretical understanding for the mass spectra of these vacua was given in ref.~\cite{Kodama:2012hu}.

The superpotential for this subsector of maximal supergravity is given by \cite{Roest:2009tt}
 \begin{align}
  W = \tfrac12 {\rm Tr} [ Q \cM ] \,,
 \end{align}
where $\cM^{ab}$ is the symmetric scalar matrix spanning the $\textrm{SL}(8) / \textrm{SO}(8)$ coset. The resulting scalar potential is given by
 \begin{align}
  V = - \tfrac38 W^2 + \tfrac14 g^{ij} \partial_i W \, \partial_j W \,.
 \end{align}
Here the scalar metric is determined by the kinetic terms, which are given by
 \begin{align}
  \cL_{\rm kinetic} =  \tfrac{1}{8} {\rm Tr} [\partial_\mu \cM \partial^\mu \cM^{-1} ] \,.
 \end{align}

Here we propose the following generalisation to include the additional phase mentioned in the introduction:
 \begin{align}
  W = \tfrac12 {\rm Tr} [ Q \cM - i P \cM^{-1}] \,. \label{superpot}
 \end{align}
Note that this superpotential is complex, and therefore one also has to adapt the definition of the scalar potential:
 \begin{align}
  V = - \tfrac38 |W|^2 + \tfrac14 g^{ij} \partial_i W \, \partial_j \overline{W} \,. \label{N=8potential}
 \end{align}
The parameters $Q_{ab}$ and $P^{ab}$ are restricted by the QC of maximal supergravity, which require their product to be pure trace. Assuming non-degenerate matrices, this implies that they are proportional. We will take
 \begin{align}
 \begin{array}{lclc}
  Q = \cos(\omega) \, \mathbb{I}_{(p,q)} & , & \qquad P = \sin(\omega) \, \mathbb{I}_{(p,q)} & ,
 \end{array}
 \end{align}
where $\mathbb{I}_{(p,q)}$ is the $\textrm{SO}(p,q)$-invariant metric.

An important subsector for our purposes will consist of the seven dilatons of the theory. For these the scalar matrix is diagonal and given by
 \begin{align}
 \label{MSL(8)}
  \cM^{ab} = \delta^{ab} {\rm exp}(\beta_{ai} \phi_i) \,,
 \end{align}
where the $\beta_{ai}$ are weights of $\textrm{SL}(8)$ and satisfy
\begin{equation}
\begin{array}{lclclc}
\sum\limits_{a}\beta_{ai}=0 & , & \sum\limits_{a}\beta_{ai}\beta_{aj}=2\delta_{ij} & , & \vec{\beta}_a\cdot\vec{\beta}_b=2\delta_{ab}-\tfrac14 & .
\end{array}
\end{equation}
In our conventions this corresponds to having $g_{ij} = \tfrac12 \delta_{ij}$. Two of these dilatons are $\textrm{SU}(4) \simeq \textrm{SO}(6)$-invariant and hence are common to the $35_{\rm v}$ as well as the $\textrm{SU}(3)$-invariant sector. The $\textrm{SU}(4)$-invariant scalar potential is obtained from \eqref{N=8potential} by identifying
 \begin{align}
  \cM = {\rm diag}( e^{\phi + \sigma}, e^{\phi - \sigma}, e^{-\phi/3}, \ldots, e^{-\phi/3} ) \,.
\end{align}
In order to make contact with the notation of section 2, one needs to identify
\begin{align*}
z = \eta \quad , \qquad \zeta_{12} = \frac{\zeta}{1 + \sqrt{1 - \zeta^{2}}} \, ,
\end{align*}
then  perform the change of variables
\begin{align*}
 \left\{ \eta \rightarrow \tanh \frac{\phi}{3} , \, \zeta \rightarrow \tanh \frac{\sigma}{2} \right\} \, .
 \end{align*}
With this change of variables, the $\textrm{SU}(4)$-invariant parts of the scalar potentials \eqref{PotFromSuperPot} and \eqref{N=8potential} coincide.

Finally, we close this appendix by investigating the extrema of the superpotential $W$. As argued in refs~\cite{DallAgata:2011aa, Kodama:2012hu}, one can restrict oneself to diagonal matrices $Q$ and $P$, as these can always be diagonalised by a basis transformation. Secondly, one can restrict oneself to the origin. Any critical point away with non-vanishing scalar expectation values in the ${\bf 35}_v$ can first of all be rotated to a basis where it is only supported by dilatons; this is the same rotation that makes $Q$ and $P$ diagonal. Subsequently, bringing the dilatonic point to the origin corresponds to rescalings of $Q$ and $P$ and hence do not affect their diagonality. Hence this situation is completely general. We will use the following form of $Q$ and $P$ appearing in the superpotential
\begin{equation}
\begin{array}{lclc}
Q_{ab}=\lambda_{a} \delta_{ab} & , & P^{ab}=\mu_{a} \delta^{ab} & ,
\end{array}
\end{equation}
where $\lambda_{a}$ and $\mu_{a}$ are arbitrary positive constants. The extremality condition for the superpotential \eqref{superpot} in this case reads
\begin{equation}
\begin{array}{lccclc}
\sum\limits_{a}^{}\lambda_{a}\beta_{ai}=0 & & \textrm{and} & & \sum\limits_{a}\mu_{a}\beta_{ai}=0 & .
\end{array}
\end{equation}
Each of the two conditions above represents a linear system of seven homogenous equations for eight real unknowns ($\left\{\lambda_{a}\right\}$ and $\left\{\mu_{a}\right\}$ respectively). Since the matrix of the coefficients, which is given by the weights of $\textrm{SL}(8)$ $\beta_{ai}$, has maximal rank, the one-dimensional space of solutions is generated by the standard AdS $\cN=8$ supersymmetric critical point with residual symmetry $\textrm{SO}(8)$. This one is given by $\lambda_{a} = \cos(\omega)$ and $\mu_{a} = \sin(\omega)$ for every $a=1,\dots,8$ and its energy is $V=-6$. It was already noted that in ref.~\cite{DallAgata:2011aa} that the CC and the mass spectrum of the maximally supersymmetric vacuum is unaffected by the phase $\omega$.

\section{The $\cN=2$  action in canonical form} \label{CanonicalN=2}

Here we will derive the $\omega \neq 0$ action (\ref{electricN=2action}) from its $\omega =0$ counterpart \cite{Bobev:2010ib}, showing in the process its compatibility with the canonical $\cN=2$ formalism. Throughout this appendix, indices $M=1, \ldots, 4$, $a=1,\ldots, 8$, $u=1, \ldots , 4$ and $i=z$ are, respectively, $\textrm{Sp}(4,\mathbb{R})$ vector indices, $\mathfrak{su}(2,1)$ adjoint indices, $\textrm{SU}(2,1)/(\textrm{SU}(2) \times \textrm{U}(1))$ curved indices and $\textrm{SU}(1,1)/\textrm{U}(1)$ curved holomorphic indices (we have denoted by  $z$ the only value that  $i$ takes on). The indices $I=0,1$ introduced in section 
\ref{sec:N=2formul} label, as usual, 'half' the vector representation of $\textrm{Sp}(4,\mathbb{R})$.

The $\omega =0$ SU(3)-invariant sector was shown in \cite{Bobev:2010ib} to be compatible with the $\cN=2$ formalism in the presence of purely electric gaugings (see \cite{Andrianopoli:1996cm} for a review). From the point of view of the $\omega=0$ electric frame of \cite{Bobev:2010ib}, the effect of a non-vanishing $\omega$ should translate into turning on $\omega$-controlled charges along both electric and magnetic gauge fields with respect to that frame. Here we will show that this expectation is indeed correct and that, moreover,  the $\omega$-dependent couplings are compatible with the dyonic formulation of $\cN=2$ supergravity. We will do this by first showing that the $\omega$-dependent scalar potential (\ref{PotFromSuperPot}) derived from either superpotential (\ref{SU(3) omega superpotential+}) or (\ref{SU(3) omega superpotential-}) conforms, in the symplectic frame of \cite{Bobev:2010ib}, to the canonical $\cN=2$ expression for the potential produced by a dyonic gauging \cite{Michelson:1996pn}  (see also \cite{deWit:2011gk}). We will then be able to read off the embedding tensor, which will enable us both to exhibit this dyonic interpretation of the $\omega \neq 0$ gaugings, and to reconstruct the full $\omega \neq 0$ $\cN=2$ action. We take this as one piece of very strong evidence that the superpotentials  (\ref{SU(3) omega superpotential+}), (\ref{SU(3) omega superpotential-}), which we originally introduced by symmetry arguments, do indeed give rise to the SU(3)-invariant sector of the $\omega \neq0$ SO(8)-gauged theories \cite{Dall'Agata:2012bb}.

The scalar potential due to a dyonic gauging in the hypermultipet sector only  \cite{Michelson:1996pn} reads, following a notation close to  \cite{deWit:2005ub}, 
\begin{eqnarray} \label{potN=2}
V= \Theta_M{}^a \Theta_N{}^b \Big[ 4 e^\cK X^M \overline{X}^N  h_{uv} k^u{}_a  \bar{k}^v{}_b
+ P^x_a P^x_b  \big( g^{i\bar{j}} f_i{}^M \bar{f}_{\bar{j}}{}^N -3e^\cK X^M \overline{X}^N \big) \Big] \, ,
\end{eqnarray}
Here,  $\Theta_M{}^a$ is the embedding tensor, and the rest of the symbols are the usual quantites related to the special K\"ahler and quaternionic-K\"ahler geometry of the manifolds (\ref{eq:ScalarMan}). We will make all these explicit below. In the symplectic frame of \cite{Bobev:2010ib}, all the dependence of (\ref{potN=2}) on the dyonically gauging parameter $\omega$ must be confined to $\Theta_M{}^a$, with all other quantities inside the square brackets being $\omega$-independent. We can thus directly import them from \cite{Bobev:2010ib}.

In terms of the complex coordinate $z$ on the unit disk,  the holomorphic sections $X^M = (X^I, F_I)$ are given by
\be
\label{holosections}
\begin{array}{lclclclc} 
X^0 = \tfrac{1}{\sqrt{2}} (1+z^3) & , & X^1=\sqrt{\tfrac{3}{2}} z(1+z) & , &  F_0 = -\tfrac{i}{\sqrt{2}} (1-z^3) & , &  F_1= i \sqrt{\tfrac{3}{2}} z(1-z) & .
\end{array}
\ee
These give rise to the K\"ahler potential
\begin{eqnarray} \label{KahlerPot}
\cK= -\log \big( i \overline{X}^M \Omega_{MN} X^N \big) = -\log (1- |z|^2)^3 \; , \qquad
\Omega = \left(\begin{array}{cc}
0 					& \mathds{1}_2  \\
- \mathds{1}_2	& 0 				  \\
\end{array} \right) \, ,
\end{eqnarray}
from where the metric  (\ref{ds2VM}) on $\textrm{SU}(1,1)/\textrm{U}(1)$ derives. The vielbeine $f_z{}^M = (f_z{}^I ,f_z{}_I ) \equiv  \partial_z (e^{\cK/2} X^M ) + \tfrac{1}{2}  e^{\cK/2} X^M \partial_z \cK$ read
\begin{eqnarray} \label{VMvielbein}
&& f_z{}^0 = \frac{3}{\sqrt{2}} \frac{z^2 + \bar z}{(1- |z|^2)^{5/2}} \, , \qquad 
f_z{}^1 = \sqrt{\frac{3}{2}} \ \frac{1+2z+2z\bar z + z^2 \bar z}{(1- |z|^2)^{5/2}} \, , \nonumber \\[5pt]
&& f_z{}_0 = \frac{3i}{\sqrt{2}} \frac{z^2 - \bar z}{(1- |z|^2)^{5/2}} \, , \qquad 
f_z{}_1 = i \sqrt{\frac{3}{2}} \ \frac{1-2z+2z\bar z - z^2 \bar z}{(1-|z|^2)^{5/2}} \, . \quad 
\end{eqnarray}

We now turn to the quaternionic-K\"ahler data entering (\ref{potN=2}), again collecting them from \cite{Bobev:2010ib}. We have already given the metric $h_{uv}$ on the hypermultiplet scalar manifold SU(2,1)/(SU(2) $\times$ U(1)) in equation (\ref{ds2HM}) of the main text. As for the gauged isometries, of the $\mathbf{8}$ Killing vectors $k_a$, $a=1, \ldots, 8$, of  $\mathfrak{su}(2,1)$ only
\begin{eqnarray} \label{KVs}
k_1 = i \zeta_1 \partial_{\zeta_1} - i \zeta_2 \partial_{\zeta_2} + \textrm{c.c.} \, , \qquad  
k_2 = \sqrt{3} i \zeta_1 \partial_{\zeta_1} + \sqrt{3} i \zeta_2 \partial_{\zeta_2} + \textrm{c.c.} \, ,
\end{eqnarray}
participate in the gauging. The corresponding momentum maps are
\begin{eqnarray} \label{Prepot1}
P_1 = -\frac{\big(1+ |\zeta_{12}|^2 \big)^2} {\big(1 - |\zeta_{12}|^2 \big)^2}
\left( \begin{array}{c}
\tfrac{1}{2} (\zeta_1 \bar{\zeta_2} + \zeta_2 \bar{\zeta_1} ) (\zeta_{12}^2 + \bar\zeta_{12}^2 ) \\
\tfrac{i}{2} (\zeta_1 \bar{\zeta_2} - \zeta_2 \bar{\zeta_1} ) (\zeta_{12}^2 + \bar\zeta_{12}^2 ) \\
 (2 + \zeta_{12}^4 + \bar\zeta_{12}^4 )(1+ |\zeta_{12}|^2 \big)^{-2} 
 \end{array} \right)
\end{eqnarray}
and
\begin{eqnarray} \label{Prepot2}
P_2 = -\sqrt{3} \frac{\big(1+ |\zeta_{12}|^2 \big)^2} {\big(1 - |\zeta_{12}|^2 \big)^2}
\left( \begin{array}{c}
\zeta_1 \bar{\zeta_2} + \zeta_2 \bar{\zeta_1}  \\
i(\zeta_1 \bar{\zeta_2} - \zeta_2 \bar{\zeta_1} )  \\
 2 ( \zeta_{12}^2 + \bar\zeta_{12}^2 )(1+ |\zeta_{12}|^2 \big)^{-2} 
 \end{array} \right) \, ,
\end{eqnarray}
where the combination $\zeta_{12}$ was introduced in (\ref{zeta12}). Note that the $P_a^x$ in (\ref{potN=2}) are just the components $x=1,2,3$ of each prepotential.

The only quantity in (\ref{potN=2}) that remains to be specified is the embedding tensor. We find that, bringing the definitions (\ref{holosections})--(\ref{Prepot2}) into (\ref{potN=2}), the latter reproduces the  potential (\ref{PotFromSuperPot}) derived from either superpotential (\ref{SU(3) omega superpotential+}) or (\ref{SU(3) omega superpotential-}), provided $\Theta_M{}^a =( \Theta_I{}^a , \Theta^{I a})$ is chosen to have non-vanishing components
\begin{equation} \label{eq:ETrealisation}
\Theta_0{}^1 = \cos \omega \, , \qquad 
\Theta^{01} = -\sin \omega \, , \qquad 
\Theta_1{}^2 = \cos \omega \, , \qquad 
\Theta^{12} = -\sin \omega \, ,
\end{equation}
This shows that our $\omega$-dependent potential (\ref{PotFromSuperPot}), with (\ref{SU(3) omega superpotential+}) or (\ref{SU(3) omega superpotential-}), can indeed be cast in canonical $\cN=2$ form. It is now easy to see that the role of $\omega$ is indeed to turn on electric and magnetic couplings with respect to the $\omega=0$ electric frame. In fact, after inserting (\ref{eq:ETrealisation}), the gauge covariant derivatives
\begin{eqnarray} \label{covDers}
D q^u = d q^u -  A^M \Theta_M{}^a k^u{}_a \, 
\end{eqnarray}
explicitly read,
\begin{eqnarray} \label{covDers2}
D q^u = d q^u -  \Big( (A^0 \cos\omega -A_0 \sin\omega ) k_1^u + (A^1 \cos\omega -A_1 \sin\omega ) k_2^u \Big)  \, .
\end{eqnarray}
We thus have a gauging along the  quaternionic-K\"ahler isometry $k_1$ (respectively, $k_2$) in (\ref{KVs}) with the graviphoton $A^0$ (respectively, the vector in the vector multiplet, $A^1$) of \cite{Bobev:2010ib}, for $\omega = n\pi$, $n =0, \pm 1, \dots$; with its magnetic dual $A_0$ (respectively, $A_1$), for $\omega = \frac{\pi}{2} +n\pi$; and with a combination of $A^0$ and $A_0$ (respectively, $A^1$ and $A_1$), for all other values of $\omega$.

Having determined the embedding tensor, we can now proceed to reconstruct the full $\cN=2$ action for the $\omega \neq0$ SU(3)-invariant sector. As we have just seen, a non-vanishing $\omega$ renders dyonic the symplectic frame of \cite{Bobev:2010ib} and, accordingly, the action in such frame would contain the magnetically charged hyperscalars dualised into tensors  \cite{Louis:2002ny,Theis:2003jj,Dall'Agata:2003yr,D'Auria:2004yi}. We could in principle use the formulae in \cite{deWit:2005ub,deWit:2011gk} to construct the action in this frame, but we instead perform a symplectic rotation into a new $\omega$-dependent electric frame where no such tensors appear. A symplectic transformation that does this job is  
\begin{align} \label{SympTrans}
\mathcal{S} = \left( \begin{matrix} A & B \\ C & D \end{matrix} \right) \, , \quad \textrm{with} \quad 
A=D= \cos \omega \ \mathds{1}_2  \quad  \textrm{and} \quad 
B=-C= -\sin \omega  \ \mathds{1}_2 \, . \quad 
\end{align}
Indeed, 
\begin{eqnarray} \label{SympCond}
\det \mathcal{S} =1 \, \qquad  \textrm{and} \qquad  \mathcal{S}^T \Omega \mathcal{S} = \Omega \, ,
\end{eqnarray}
where $ \Omega$ is the symplectic form in (\ref{KahlerPot}), and thus $\mathcal{S} $ is in Sp$(4,\mathbb{R})$. In the new frame, the gauge fields are $A^{\prime M} = \mathcal{S}^M{}_N A^N$, the embedding tensor is purely electric, $\Theta_M{}^{\prime a} = ( \Theta_I{}^{\prime a} = \mathds{1}_2, \Theta^{\prime Ia} = 0)$, and the covariant derivatives (\ref{covDers2}) reduce to (\ref{covDersElectric}), with the Killing vectors in (\ref{KVsElectric}) and (\ref{KVs}) related as $k^\prime_I =  \Theta_I{}^{\prime a}  k_a$. Finally, the gauge kinetic matrix $\cN_{IJ}$ of \cite{Bobev:2010ib} transforms as
\begin{align} 
\cN^\prime  = (C + D \cN)(A + B \cN)^{-1} \, 
\end{align}
into the new frame, yielding the result (\ref{gaugeKinPrime}) brought to the main text. This concludes the proof that our action (\ref{electricN=2action}) is indeed compatible with the (electric frame) formalism of $\cN=2$ gauged supergravity \cite{Andrianopoli:1996cm}.

We would like to conclude by emphasising that the scalar kinetic terms in (\ref{electricN=2action}) remain unaffected by the presence of $\omega$. Indeed, although the sections in the primed frame, $X^{\prime M} = \mathcal{S}^M{}_N X^N$, where $X^N$ are given in (\ref{holosections}), do aquire an $\omega$-dependence, this drops out from the K\"ahler potential (\ref{KahlerPot}), which remains invariant by the symplectic property (\ref{SympCond}).

\section{$\cN=2$ vacua from the canonical formalism}\label{app:N=2_vacua}

Here we will use the results of \cite{Hristov:2009uj, Louis:2012ux} to analytically find the location of the critical points that preserve the full $\cN=2$ supersymmetry of the SU(3)-invariant sector action. It is straightforward to check that this sector does not support $\cN=2$ Minkowski, AdS$_2 \times S^2$ or pp-wave spacetimes, as the Killing prepotentials (\ref{Prepot1}), (\ref{Prepot2}) are everywhere non-vanishing \cite{Hristov:2009uj}. Focusing thus on $\cN=2$ AdS vacua and particularising to hypermultiplet gaugings, the (symplectically completed) conditions for $\cN=2$ supersymmetry read \cite{Hristov:2009uj, Louis:2012ux}
\begin{eqnarray} \label{N=2Consgen}
X^M \Theta_M{}^ a k^u{}_a = 0 \, , \qquad
\epsilon_{xyz} X^M \bar{X}^N \Theta_M{}^a \Theta_N{}^b P_a^y P_b^z = 0   \, , \qquad
f_z{}^M \Theta_M{}^ a P_a^ x = 0 \, .
\end{eqnarray}
Inserting the explicit expressions for the special geometry data and the embedding tensor given in section \ref{sec:N=2formul}, we find that these conditions are equivalent to
\begin{eqnarray} \label{N=2cond1}
&& \textrm{either} 
\quad \zeta_1 =0 \qquad 
\textrm{or} \qquad 
z^3 +3e^{2i\omega} z^2 +3z+e^{2i \omega} = 0  \, , 
\end{eqnarray}
and
\begin{eqnarray} \label{N=2cond2}
&& \textrm{either} 
\quad \zeta_2 =0 \qquad 
\textrm{or} \qquad 
z^3 -3e^{2i\omega} z^2 -3z+e^{2i \omega} = 0  \, , 
\end{eqnarray}
and 
\begin{eqnarray} \label{N=2cond3}
&& \textrm{either} 
\quad \bar\zeta_1 \zeta_2  =0 \qquad
 \textrm{or} \qquad 
(1+z \bar z) \big(z e^{-2i\omega} - \bar z e^{2i\omega} \big) + z^2 -\bar z^2 =0 \, ,
\end{eqnarray}
and
\begin{eqnarray} \label{N=2cond4}
\textrm{either} \;\;
 \bar\zeta_1 \zeta_2  =0 \;\;
 \textrm{or} \; \;
(\zeta_{12}^2  + \bar \zeta_{12}^ 2 ) (z^2+ \bar z e^{2i\omega} ) + 2 ( z^2 \bar z e^{2i\omega}  + 2 z  \bar z + 2 z e^{2i\omega} + 1)  =0 \, , \nn\\
\end{eqnarray}
and
\begin{eqnarray} \label{N=2cond5}
&& (2+ \zeta_{12}^4 + \bar \zeta_{12}^4 ) (z^2+ \bar z e^{2i\omega} ) + 2 (\zeta_{12}^2  + \bar \zeta_{12}^ 2 ) ( z^2 \bar z e^{2i\omega}  + 2 z  \bar z + 2 z e^{2i\omega} + 1)   = 0 \  . \nn\\
\end{eqnarray}
Note that only the last relation, (\ref{N=2cond5}), does not involve a conditional clause. It is easy to check that, evaluated on the conditions (\ref{N=2cond1})--(\ref{N=2cond5}), both ${\cal W}_+ $ and ${\cal W}_- $  in (\ref{SU(3) omega superpotential+}) and (\ref{SU(3) omega superpotential-}) reduce to the same expression,
\begin{eqnarray} \label{eq:SuperPotN=2}
{\cal W} \equiv {\cal W}_+ = {\cal W}_- = \frac{z^3 +e^{2i\omega}}{(1-|z|^2)^{3/2}} \, ,
\end{eqnarray}
and that the resulting superpotential $|{\cal W}|$ is indeed extremised under these conditions. Furthermore, defining ${\cal W}^x = e^{{\cal K}/2} X^M \Theta_M{}^a P_a^x$, $x=1,2,3$, it can be verified that ${\cal W}^3 = -\frac{1}{\sqrt{2}} e^{-i\omega} ( {\cal W}_+ + {\cal W}_- )$ and that, under (\ref{N=2cond1})--(\ref{N=2cond5}), ${\cal W}^1 ={\cal W}^2 =0 $. This thus provides a crosscheck that the cosmological constant 
\begin{eqnarray} \label{eq:PotN=2fromSuperPot}
V_0 = -6 |{\cal W}|^2 \, 
\end{eqnarray}
with ${\cal W}$ given in (\ref{eq:SuperPotN=2}) for an ${\cal N} =2$ critical point, agrees when calculated in the superpotential, (\ref{PotFromSuperPot}), and canonical, (\ref{potN=2}), (\ref{N=2Consgen}) formalisms. More importantly, (\ref{eq:SuperPotN=2})  provides a useful simplification to explicitly evaluate the cosmological constant (\ref{eq:PotN=2fromSuperPot}) at an ${\cal N} =2$ point.

The origin of the scalar manifold in these coordinates, $z=\zeta_1 = \zeta_2 = 0$, solves all the requirements (\ref{N=2cond1})--(\ref{N=2cond5}) for $\cN=2$ supersymmetry. This corresponds to the SO(8) point, which indeed is $\cN=2$ within the SU(3)-invariant truncation: its supersymmetry is only enhanced to $\cN =8$ in the full maximal theory. To systematically search for all other possible $\cN=2$ points, we only need to consider three cases, $\zeta_2=0$, $\zeta_1=0$ and $\bar\zeta_1 \zeta_2 \neq 0$, in (\ref{N=2cond1})--(\ref{N=2cond5}).

Let us first set $\zeta_2= 0$. In this case, we only need to impose the second equation in (\ref{N=2cond1}) and equation (\ref{N=2cond5}). These allow us to solve for $z$ and $\zeta_{12}$, respectively. We find that the vanishing locus of the cubic in (\ref{N=2cond1}) is most easily studied in the upper-half plane, where this equation is mapped, via (\ref{DiskToUHF}), into the much simpler
\begin{equation}  \label{N=2cond1UHP}
\cos\omega - S^3 \sin\omega =0 \, .
\end{equation}
Equation (\ref{N=2cond1UHP}) is now trivial to solve: it has no solutions for $\omega = n \pi$, $n=0, \pm 1, \dots$; otherwise, it has three roots,
\begin{equation} \label{Roots1UHP}
S_k = e^{\frac{2ik\pi}{3}} (\cot\omega )^{1/3}  \, , \qquad
k=1,2,3 \, ,
\end{equation}
lying at the vertices of an equilateral triangle centered at the origin of the $S$ complex plane, except for $\omega = \frac{\pi}{2} + n \pi$, $n=0, \pm 1, \dots$, where the triangle degenerates into a triple root. In (\ref{Roots1UHP}) we are taking $(\cot\omega)^{1/3}$ to be real and with the same sign than $\cot\omega$: positive and negative for $\omega \in (0,\frac{\pi}{2})$ and $\omega \in (\frac{\pi}{2},\pi)$, respectively. Accordingly, we only have one physical solution, lying on the strict upper half plane, for each $\omega$: either $S_1$ or $S_2$ in each interval of $\omega$. The third root, $S_3$, lies on the real axis and is thus unphysical for all $\omega$. Mapping back onto the unit disk via (\ref{DiskToUHF}), we have a unique physical solution to the cubic in (\ref{N=2cond1}), which can be written as
\begin{eqnarray} \label{eq1Sol}
z = z_0 (\omega), \quad  \textrm{for $\omega \in (0,\tfrac{\pi}{2})$} \, , \qquad 
z= {\bar z}_0 (\omega) , \quad  \textrm{for $\omega \in (\tfrac{\pi}{2},\pi)$} \, ,
\end{eqnarray}
with $z_0 (\omega)$ given in (\ref{defsN=2}). Although in (\ref{defsN=2}) we have followed the same sign convention for $(\tan\omega)^{1/3}$ mentioned above, we have used absolute value in order to avoid any confusion. Finally, we can insert these in (\ref{N=2cond5}) and solve for $\zeta_{12}$. Note that, for $\zeta_2 =0$, it is $\bar \zeta_{12} = \zeta_{12}$, and thus (\ref{N=2cond5}) is quadratic in the real variable $\zeta_{12}^2$. Solving thus for $\zeta_{12} = |\zeta_{12}|$ we find the expression in (\ref{branch2N=2}), (\ref{defsN=2}), noting here that these expressions for $\zeta_{12}$ are actually valid for all $\omega \in (0,\pi)$. Finally, some manipulations allow us to write the cosmological constant 
(\ref{eq:PotN=2fromSuperPot}), (\ref{eq:SuperPotN=2}) as $V_0 = P(\omega)$, where we have given $P(\omega)$ in (\ref{eq:PotN=2}). This solution corresponds to the new branch (\ref{branch2N=2}) of ${\cal N} = 2$ critical points, evaluated here for all $\omega \in (0,\pi)$.

Turning now to the case $\zeta_1=0$, only the second equation in (\ref{N=2cond2}) and equation (\ref{N=2cond5}) need to be imposed. We now use symmetry to solve for the former: if $(z, \omega)$ solve the cubic in (\ref{N=2cond1}) that we have just studied, then $(z^\prime, \omega^\prime ) = (-i\bar z, -\omega +\frac{\pi}{4})$ solve the cubic in (\ref{N=2cond2}). Applying this transformation to (\ref{eq1Sol}), (\ref{defsN=2}), we find that the unique physical solution to the cubic in (\ref{N=2cond2}) is
\begin{eqnarray} \label{eq2Sol}
z = -i \ \bar{z}_0 \big(\omega - \tfrac{\pi}{4} \big) , \quad  \textrm{for $\omega \in (-\tfrac{\pi}{4},\tfrac{\pi}{4})$} \, , \qquad 
z = -i \ z_0 \big(\omega - \tfrac{\pi}{4} \big) , \quad  \textrm{for $\omega \in (\tfrac{\pi}{4}, \tfrac{3\pi}{4})$} \, .
\end{eqnarray}
These can again be plugged in (\ref{N=2cond5}) to solve for $\zeta_{12}$. Note that, for $\zeta_1 =0$, it is $\bar \zeta_{12} = -\zeta_{12}$, so (\ref{N=2cond5}) is again quadratic in the real variable $\zeta_{12}^2$. Solving thus for $\zeta_{12} = i |\zeta_{12}|$ we find the expression in (\ref{branch1N=2}), (\ref{defsN=2}), noting again that these expressions  are  valid for all $\omega$. Finally, some manipulations allow us to write the cosmological constant 
(\ref{eq:PotN=2fromSuperPot}), (\ref{eq:SuperPotN=2}) as $V_0 = P(\omega -\tfrac{\pi}{4})$ in this case. This solution, corresponding to the branch containing the $\omega =0$ point, corresponds to the branch (\ref{branch1N=2}) of ${\cal N} = 2$ critical points. Here we have given it for all $\omega$.

We finally consider the case $\bar\zeta_1 \zeta_2 \neq 0$. It is easy to see that there are no solutions in this case. Indeed, the cubics in (\ref{N=2cond1}) and (\ref{N=2cond2}) need now to be simultaneously imposed, and their solutions, (\ref{eq1Sol}) and  (\ref{eq2Sol}),  do not overlap at any fixed $\omega$. An alternative and quicker way to see that there are no physical solutions in this case is to sum the cubics in (\ref{N=2cond1}) and (\ref{N=2cond2}): the resulting cubic is straighforwardly seen to have solutions only at the unphysical boundary of the disk.

\section{Minkowski vacua for other gaugings of the $\cN=8$ theory}\label{app:other_gaugings}



We will present four families of Minkowski critical points, \textit{i.e.} $V_{0}=0$,  with different residual symmetry groups and preserving different amounts of supersymmetry. All these follow from the general analysis of $\textrm{SU}(3)$-invariant embedding tensors\footnote{We have found no de Sitter critical points with an $\textrm{SU}(3)$ residual symmetry group and with an arbitrary amount of preserved supersymmetry.} as outlined in section 4. However, none of these families is compatible with an $\textrm{SO}(8)$ gauging for any value of the parameters since no Minkowski solutions were found in section 3 by using the superpotential approach.

\subsubsection*{$\textrm{U}(4)$-invariant vacua with $\cN = 6$ supersymmetry}

This family of Minkowski solutions involves the following configuration for the embedding tensor parameters
\be
\begin{array}{lclclc}
\lambda_{1} =  \Lambda \, e^{-i3\theta} & , & \mu_{5} = -\mu_{6} = -\Lambda \, e^{i\theta} & ,
\end{array}
\ee
where $\Lambda$ is an arbitrary (real) scaling parameter and all the other embedding tensor parameters are vanishing. The associated embedding tensor configuration happens to be $\textrm{U}(4)$-invariant. 
Please note that the $\textrm{SU}(4)$ factor sitting inside this $\textrm{U}(4)=\textrm{SU}(4) \times \textrm{U}(1)$ symmetry is realised in a different way from the one of the $\textrm{SU}(4)$ truncation presented in section~\ref{sec:N=8_spectra}.
This residual symmetry should be rather interpreted as $\textrm{SO}(2) \times \textrm{SO}(6)$ acting in the following way on the $\textrm{SU}(8)$ indices
\be
\begin{array}{cccc}
\textrm{SU}(8) & \longrightarrow & \textrm{SO}(2) \times \textrm{SO}(6) & , \\[2mm]
 \mathcal{I} & \longrightarrow & (\underbrace{\,1,\,\,\hat{1}\,}_{\textrm{SO}(2) \textrm{ doublet}}, \,\underbrace{\, a,\,\,\hat{a}\,}_{\textbf{6}\textrm{ of SO}(6)}) & ,
\end{array}
\ee
the $\textrm{SO}(6)$ factor being the R-symmetry of the $\cN = 6$ theory.

The scalar mass spectrum is very simple for this family of critical points and is given by
\be
\begin{array}{lclclclclc}
m^2 & = &  0 \quad (\times 42) & , & 2 \, \Lambda^{2} \quad (\times 28) & , 
\end{array}
\ee
whereas the vector masses are
\be
\begin{array}{lclclclclc}
m^2 & = &  0 \quad (\times 44) & , & 2 \, \Lambda^{2} \quad (\times 12) & .
\end{array}
\ee
Apart from the $28$ unphysical vectors, there are $16$ physical massless vectors associated to the $\textrm{U}(4)$ residual symmetry group.

\subsubsection*{$(\textrm{SU}(3) \times \textrm{U}(1)^{(1)} \times \textrm{U}(1)^{(2)})$-invariant vacua with $\cN = 2$ supersymmetry}

This family of Minkowski solutions is compatible with an embedding tensor configuration given by the non-vanishing components
\be
\begin{array}{lclclc}
\lambda_{2} = \pm \frac{\Lambda}{2} \, e^{-i 7 \theta} & , & \mu_{7} = \mu_{8} = \Lambda \, e^{i 3 \theta} & , & \mu_{9} = -\mu_{10} = \mp \frac{\Lambda}{2} \, e^{i \theta} & .
\end{array}
\ee
Using the charge assignments in table~\ref{Table:U(1)U}, it is easy to check that these components are invariant under both $\textrm{U}(1)^{(1)}$ and $\textrm{U}(1)^{(2)}$ simultaneously. 

The scalar mass spectrum at this family of critical points reads
\be
\begin{array}{lclclclclc}
m^2 & = &  0 \quad (\times 36) & , & \dfrac{\Lambda^{2}}{2} \quad (\times 24) & , & 2 \, \Lambda^{2} \quad (\times 6) & , & \dfrac{9 \, \Lambda^{2}}{2} \quad (\times 4) & ,
\end{array}
\ee
whereas the vector masses are given by
\be
\begin{array}{lclclclclc}
m^2 & = &  0 \quad (\times 38) & , & \dfrac{\Lambda^{2}}{2} \quad (\times 12) & , & 2 \, \Lambda^{2} \quad (\times 6)  & .
\end{array}
\ee
Out of the 38 massless vectors, there are 10 which are physical and are associated to the residual symmetry group.

\subsubsection*{$(\textrm{SU}(4) \times \textrm{U}(1)_{S})$-invariant vacua with $\cN = 0$ supersymmetry}

In order to describe this family of non-supersymmetric Minkowski solutions it is convenient to go to the intermediate $\textrm{SU}(4)$ truncation involving the embedding tensor components in (\ref{SU4A1}) and (\ref{SU4A2}). In terms of these, the embedding tensor parameter are given by
\be
\begin{array}{lclclc}
\gamma = - \frac{\Lambda}{2} \, e^{-i3\theta} & , & \delta_{1} = \delta_{3} = \Lambda \, e^{i\theta} & .
\end{array}
\ee
One can verify that these components are invariant under $\textrm{U}(1)_{S} \subset \textrm{SL}(2)_{S}$ rotations transforming indices $i$ and $\hat{i}$ with charges $-1$ and $+1$, respectively. 

The scalar masses at this family of critical points are given by
\be
\begin{array}{lclclclclc}
m^2 & = &  0 \quad (\times 48) & , & 2 \, \Lambda^{2} \quad (\times 20) & , & 8 \, \Lambda^{2} \quad (\times 2)  & ,
\end{array}
\ee
and coincides with those of the $\textrm{SO}(2,6)$ gauging presented in ref.\cite{DallAgata:2011aa} after setting $\Lambda^2=1/4$. The vector masses read
\be
\begin{array}{lclclclclc}
m^2 & = &  0 \quad (\times 44) & , & 2 \, \Lambda^{2} \quad (\times 12)   & ,
\end{array}
\ee
so there are 16 physical massless vectors associated to the $\textrm{U}(4)$ residual symmetry.

\subsubsection*{Interpolating between Minkowski vacua}

The above three sets of Minkowski solutions preserving different amount of supersymmetry can be unified into a bigger $2$ (real) $+$ $1$ (phase) parameter family of Minkowski solutions. This is specified in terms of the following real $\textrm{SU}(3)$-invariant embedding tensor parameters
\be
\begin{array}{lclc}
\lambda_{1} = -\mu_{5} = \mu_{6} = \Lambda_{1} \, \cos(\theta) &  , &  \lambda_{3} =- \lambda_{4}=\mu_{15} =  \mu_{16} = - \Lambda_{1} \, \sin(\theta)  & , \\[2mm]
2 \lambda_{2} =\mu_{7} = \mu_{8} =-2 \mu_{9} = 2 \mu_{10} = - 2 \Lambda_{2}  & .
\end{array}
\ee
The amount of supersymmetry preserved by the solutions in this family depends on the values of the $\Lambda_{1,2}$ parameters :  $\,i)\,$ $\cN=6$ for $\Lambda_{1}\neq 0$ and $\Lambda_{2} = 0\,\,$  $ii)\,$ $\cN=2$ for $\Lambda_{1}=0$ and $\Lambda_{2}\neq 0\,\,$  $iii)\,$ $\cN=0$ whenever $\Lambda_{1} \, \Lambda_{2} \neq 0$ .

The spectra of scalar and vector masses are also $\Lambda_{1,2}$ dependent quantities. The former is given by
\be
\begin{array}{lclclclclc}
m^2 & = &  2 \,( \Lambda_{1} \pm \Lambda_{2})^{2} \quad (\times 12)  & , & 8 \, \Lambda_{2}^{2} \quad (\times 6) & , & 0 \quad (\times 36) & ,\\[2mm]
&  & 2 \,( \Lambda_{1} \pm 3 \, \Lambda_{2})^{2} \quad (\times 2)  & ,
\end{array}
\ee
whereas the latter reads
\be
\begin{array}{lclclclclc}
m^2 & = &  0 \quad (\times 38) & , & 8 \, \Lambda_{2}^{2} \quad (\times 6)   & , & 2 \,( \Lambda_{1} \pm \Lambda_{2})^{2} \quad (\times 6) & .
\end{array}
\ee
By inspection of the above mass spectra, one observes some special limits :
\begin{itemize}

\item $\Lambda_{1}=0$ : In this case, the scalar and vector mass spectra coincide with those of the previous $\cN=2$ family of Minkowski solutions after the identification $\Lambda_{2}=\Lambda/2 $.

\item $\Lambda_{2}=0$ : In this case, the scalar and vector mass spectra coincide with those of the previous $\cN=6$ family of Minkowski solutions after the identification $\Lambda_{1}=\Lambda $.

\item $\Lambda_{1}=\pm \Lambda_{2}$ : In this case, the scalar and vector mass spectra coincide with those of the previous $\cN=0$ family of Minkowski solutions after the identification $\Lambda_{1}=\pm \Lambda_{2}=\Lambda/2 $.

\item $\Lambda_{1}=\pm 3\, \Lambda_{2}$ : This case does not reduce to any of the previous $\cN=0$ families of Minkowski solutions. Moreover, it has an enhancement of massless scalars with respect to the generic case not being associated to the presence of additional massive vectors, \textit{i.e.} to further symmetry breaking.

\end{itemize}
A remarkable feature of the generic case is that it does not contain tachyons even though supersymmetry is completely broken.

\bibliography{SU3ref}
\bibliographystyle{utphys}

\end{document}